\documentclass{aa}  

\usepackage{graphicx}
\usepackage{txfonts}
\newcommand{\cloudy}{{\fontfamily{qcr}\selectfont CLOUDY}}
\usepackage{hyperref}
\begin{document}

   \title{Broad-line region in active galactic nuclei:  Dusty or dustless?}

   \subtitle{}

   \author{Ashwani Pandey\inst{1}
          \and
          Bo\.zena Czerny\inst{1}
          \and
          Swayamtrupta Panda\inst{2}\fnmsep\thanks{CNPq Fellow}
          \and
          Raj Prince \inst{1}
          \and 
          Vikram Kumar Jaiswal \inst{1}
          \and
          Mary Loli Martinez–Aldama\inst{3}
          \and
          Michal Zaja\v{c}ek\inst{4}
          \and
          Marzena \'Sniegowska\inst{5}
          }

\institute{
Center for Theoretical Physics, Polish Academy of Sciences, Al. Lotnik\'ow 32/46, 02-668 Warsaw, Poland
    \and
Laborat\'orio Nacional de Astrof\'isica, MCTI, Rua dos Estados Unidos 154, Bairro das Na\c c\~oes. CEP 37504-364, Itajub\'a, MG, Brazil
    \and
Astronomy Department, Universidad de Concepción, Casilla 160-C, Concepción, 4030000, Chile 
\and
Department of Theoretical Physics and Astrophysics, Faculty of Science, Masaryk University, Kotl\'a\v{r}ska 2, 611 37 Brno, Czech Republic
\and
School of Physics and Astronomy, Tel Aviv University, Tel Aviv 69978, Israel
}

   \date{Received XXX XX, 2023; accepted XXX XX, XXXX}

% \abstract{}{}{}{}{} 
% 5 {} token are mandatory
 
  \abstract
   {Dust in active galactic nuclei is clearly present right outside the broad-line region (BLR) in the form of a dusty molecular torus. However, some models of the BLR predict that dust may also exist within the BLR.}
   {We study the reprocessing of radiation by the BLR with the aim of observing how the presence of dust affects the reprocessed continuum and the line properties.}
   {We calculated a range of models using the \cloudy \ photoionisation code for dusty and dustless plasma. We paid particular attention to the well-studied object NGC 5548, and we compared the line equivalent width predictions with the data from observations for this object.}
   {We obtained a rough agreement between the expected equivalent widths of the H$\beta$ and Mg II lines and the observed values for NGC 5548 for the line distances implied by the time-delay measurement (for H$\beta)$ and the radius-luminosity relation (for Mg II) when the medium is dusty. We found the incident radiation to be consistent with the radiation seen by the observer, so no shielding between the inner disc and the BLR is required. High ionisation lines such as He II, however, clearly form in the inner dustless region. When the additional absorber is present, the H$\beta$ emitting region moves closer to the dustless part of the accretion disc surface.}
   {}


   \maketitle
%
%-------------------------------------------------------------------

\section{Introduction}

Dust is one of the key constituents of most active galactic nuclei (AGN). Its presence has been well established by spectroscopic observations and direct interferometric mapping of nearby AGN \citep[see e.g.][for short recent reviews]{ramos_almeida_2017, Lyu_Rieke_2021, czerny2023}, but the key understanding of the role of dust in AGN only came with the realisation that dust, predominantly present in the flat form of a torus, could be responsible for the classification of AGN as type 1 sources, where the inner nucleus is unshielded by dust and visible, and type 2 sources, where the inner nucleus is shielded from the observer by a high inclination angle with respect to the symmetry axis \citep{antonucci1985}. Models of the dusty torus imply the location of the dust as  being at a fraction of a parsec distance from the central supermassive black hole, depending on the bolometric luminosity of the source, as the inner radius is set by the dust sublimation temperature \citep{barvainis1987, Netzer_Laor_1993,nenkova2008}. By now, the presence of dust has been well studied observationally \citep[see e.g.][]{Honig_etal_2010, Kishimoto_etal_2011, Burtscher_etal_2013, Tristram_etal_2014, Hickox_Alexander_2018, Stalevski_etal_2019, Leftley_etal_2021, GRAVITY_2023}.

Spectroscopically, the infrared emission was first identified as the 3 $\mu$m bump in AGN spectra \citep{neugebauer1979,barvainis1987}, and ground-based and satellite observations later allowed for detailed study of the dust's characteristic features \citep[][]{alonso_herrero2003,osterbrock_book_2006,kirkpatrick2012,alonso_herrero_2012,garcia_bernete2022,donnan2023}. Reverberation mapping (RM) studies have provided an insight into the dust location through the measurement of the time delays in the IR bands \citep[see e.g.][]{Pozo-Nunez_etal_2015, Schnulle_etal_2015, Lyu_etal_2019, Shablovinskaya_etal_2020, Sobrino_Figaredo_etal_2020,  Yang_etal_2020, Lyu_Rieke_2021, Guise_etal_2022}. Monitoring of Fairall 9 with the International Ultraviolet Explorer (IUE) and ground-based IR South African Astronomical Observatory (SAAO) telescope \citep{clavel1989} showed the location of the dust ($\sim$1-year time delay) as being right outside of the CIV emitting region ($\sim$150-day delay) based on time-delay measurements relative to the UV continuum. A systematic study of the IR delay showed that the dust is located further from the black hole than the broad-line region (BLR; \cite{koshida2014}) by a factor of five. The most recent study of the mid-IR delays by \citet{chen2023} reports the ratio of the K-band ($\sim$2.2 $\mu$m), W1 ($\sim$3.4 $\mu$m), and W2 ($\sim$4.6 $\mu$m) band delay to the BLR delay as being 6.2, 9.2, and 11.2., respectively 

Direct interferometric mapping of the dust has revealed that the geometry of the dust distribution is in fact much more complicated than envisioned by the early models of a continuous or clumpy torus \citep[][]{nenkova2008}. Apart from the dust confined to the equatorial plane (i.e. torus), there is a clear presence of polar dust \citep{Honig_etal_2012, seba2013, Tristram_etal_2014, Asmus_etal_2016, Lopez-Gonzaga_etal_2016, seba2019, garcia_bernete2022, Cerqueira-Campos_etal_2023}, and some models now incorporate this element \citep[e.g.][]{siebenmorgen2015,seba2017}. The same picture was claimed on the basis of RM studies of NGC 4151 in several near-IR bands, that is, the authors of such studies have claimed that some of the signals must have come from the polar regions \citep{Lyu_Rieke_2021, Cerqueira-Campos_etal_2023}.

However, the dust RM has proven that dust can be present much closer than implied by the dust sublimation temperature \citep[e.g. by a factor of three][]{kishimoto2007}. Moreover, \citet{temple2021} noticed an interesting relation between the CIV outflow and the 2$\mu$m bump that suggests a link between the properties of the BLR and the IR-emitting dusty regions in quasars. Thus, although most of the dust is certainly located outside the BLR, some of it could actually be inside, if properly shielded from the nuclear emission.

Historically, the presence of dust inside BLR clouds was considered in a number of publications \citep[e.g.][]{martin1980,rudy1982}. The need for dust in the BLR was claimed on the basis of the measured ratio of H$\alpha$ to H$\beta$ \citep{osterbrock1981}. This ratio, from recombination calculations, was expected to be equal to 2.85 for the adopted local density of $10^4$ cm$^{-3}$ and temperature $10^4$ K \citep{broclehurst1971}. However, the measured values of the H$\alpha$-H$\beta$ ratio were frequently different, and the difference was attributed to dust co-existing with the BLR clouds \citep[see e.g.][]{Goodrich_1995}. Later, the issue of the presence of dust inside the BLR was largely forgotten since the idea of a dusty torus outside the BLR emerged. 

However, two theoretically motivated models of the BLR have been proposed in recent years that are based on dust presence in the BLR. These models possibly apply to the low ionisation line (LIL) part of the BLR, according to the distinction introduced by \citet{suzy1988}. The outer disc surface, if not flaring too strongly, is not irradiated strongly enough for the dust to be destroyed even at distances of the BLR (i.e. much closer than the torus).

The first of the two scenarios is a dynamical model of the BLR clouds, which are launched from the outer parts of the accretion disc under the radiation pressure acting on dust. \citet{czhr2011} proposed it as a failed radiatively accelerated dusty outflow \citep[FRADO; see ][]{czerny2015,czerny2017}, but recent 2.5D modelling with realistic opacities have shown that, for a high Eddington rate and/or high metal content, apart from the failed wind, the model also produces an outflowing stream of material \citep{naddaf2021,naddaf2022}. The model is not only consistent with recent dust mapping trends, but it also roughly consistently predicts the properties of the broad absorption line (BAL) quasars \citep{naddaf_BAL_2022}.  In this model, dusty and dustless clouds can coexist since their dust content depends on their orbit.

The second scenario involves a static model of \citet{baskin2018}. In this model, the dust inside the disc affects its structure, and the disc remains in hydrostatic equilibrium (for most radii, where the model can be calculated). Also in this model, part of the disc remains dustless, but the shielded part remains dusty.

The issue of the H$\alpha$ to H$\beta$ line ratio and the presence of dust in this context was also revived by \citet{gaskell2017}. The presence of dust in (some) BLR clouds can affect the BLR emissivity, and this can be important in several contexts. Dust modifies gas line emissivity \citep[e.g.][]{Netzer_Laor_1993,adhikari2016}, which might be important for the determination of the BLR covering factor \citep[see e.g.][]{Maiolino_etal_2001, Mor_Trakhtenbrot_2011,baskin2018}. Dust can also modify the continuum emission from the BLR, and this, in turn, is important for the potential measurements of the intrinsic continuum time delays, which require subtraction of the BLR contamination \citep[see][]{netzr2022,vikram2023, Pozo-Nunez_etal_2023}. 

Therefore, in this paper, we systematically compare the emissivity of the BLR with and without the dust in the BLR clouds. We mostly follow the excellent paper of \cite{2019MNRAS.489.5284K}, who studied dustless clouds, and for easy comparison, we adopt a similar parameter grid in our simulations. We specifically concentrate on comparing our results for dusty and dustless BLR for the well-studied object NGC 5548. Our general method is given in Section~\ref{sect:method}. The results from the reprocessing of the BLR continuum and emission lines are presented in Section~\ref{sect:results}. In Section~\ref{sect:with_delay}, we combine the results for the reprocessing with the actual effective location of Mg II and H$\beta$ known from the time delays, and we show that this region is most likely dusty. Several investigations of NGC 5548 have reported the possibility of the presence of an equatorial obscurer between the nucleus and the BLR \citep[e.g.][]{holidays2019, dovciak2022}, we address this issue in  Section~\ref{sect:filtering},  and Section~\ref{sect:discussion} contains the discussion. We summarise our findings from this study in Section~\ref{sect:conclusion}.

\begin{figure*}
\includegraphics[width=9cm, height=9cm]{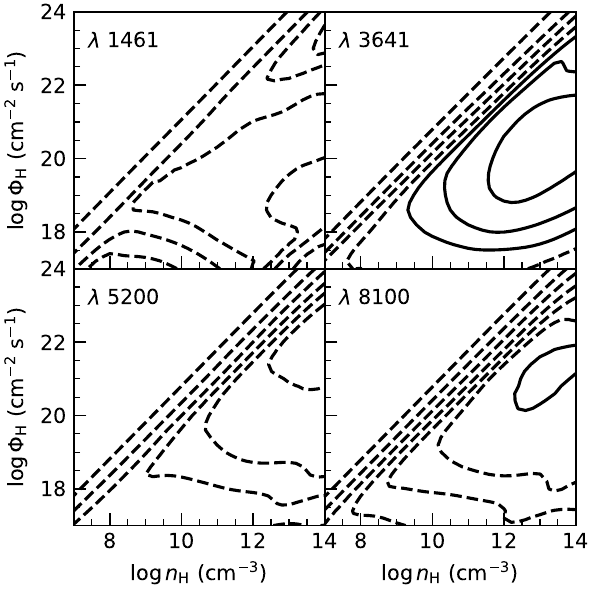}\hfill \includegraphics[width=9cm, height=9cm]{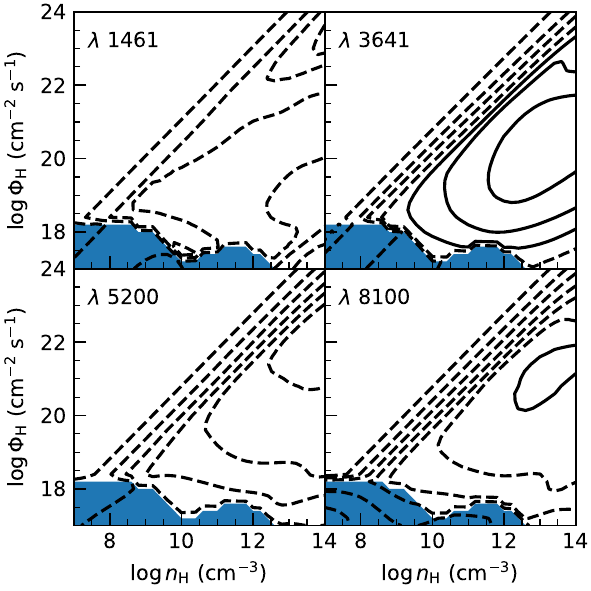}
\caption{\label{fig:contours}Logarithm contours of the ratio of diffuse continuum emission at four wavelength bands to the incident continuum emission at 1215 $\AA$ in the $\log n_{\rm H}- \log \Phi_{\rm H}$ plane for the dustless (left panel) and dusty (right panel) models. The shaded region in the right panel represents the solutions for the dusty cloud for a sublimation temperature of 2000 K.}
\end{figure*}

\begin{figure}
\includegraphics[height=6cm,width=9cm]{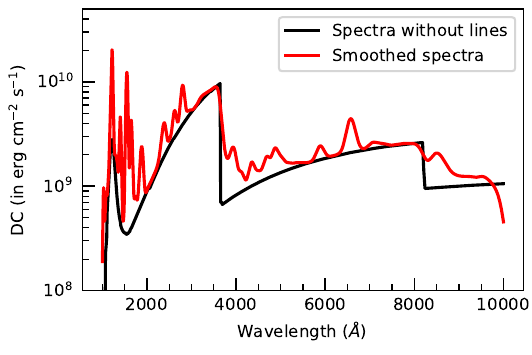}
\caption{\label{fig:spectra}Example of the BLR spectrum for only the continuum and the continuum with lines coming from \cloudy \ modelling smeared with the velocity appropriate for NGC 5548 ($\sim$4000 km/s for H$\beta$; \cite{pei2017}). $\log \Phi_{\rm H} (\rm cm^{-2} s^{-1}) = 20$, log n$_{\rm H}$ (cm$^{-3}$) = 12. }
\end{figure}

\section{Methods}
\label{sect:method}

We generated a grid of models for individual clouds in a plane-parallel approximation using the photoionisation code \cloudy, version 22.01 \citep{2017RMxAA..53..385F}.  In computations, we assumed a constant gas density case, as in \citep{2019MNRAS.489.5284K}, for a natural comparison of the dust's effect on cloud emissivity.

We parameterised the solutions with the local density of the cloud, $n_{\rm H}$, and the incident radiation flux, $\Phi_{\rm H}$. We adopted the parameter range from \cite{2019MNRAS.489.5284K}, namely, 7$\leq \log n_{\rm H}\, (\rm cm^{-3}) \leq$ 14 and 17 $\leq \log \Phi_{\rm H}\,(\rm cm^{-2} s^{-1}) \leq$ 24, with a step size of 0.2 in the logarithm scale of each parameter, which yielded a total of 1296 models. We adopted a fixed hydrogen column density of $\log N_{\rm H}$ (cm$^{-2}$) = 23  for  \cloudy \ simulations, as recommended by many authors \citep[e.g.][]{netzer2010, DuShou2023}. We assumed solar abundance for the BLR clouds.

For the incident radiation, we adopted the continuum spectral energy distribution (SED) for well-studied AGN NGC 5548 from \cite{2015A&A...575A..22M}. The SED is available in the \cloudy \ database in a file ``NGC5548.sed". 

For a comparison of the dust effect, we simulated both dusty and dustless clouds. We employed graphite dust grains with a size distribution similar to that of Orion for the computation of dusty BLR models. The composition and physics of the grains are described in
\cite{1991ApJ...374..580B},  \cite{2004MNRAS.350.1330V}, and  \cite{2006ApJ...645.1188W}. The dust cannot survive at very high temperatures and is destroyed in a very short timescale. In our calculations, we set the dust sublimation temperature to a somewhat arbitrary value of 2000 K \citep{baskin2018}. %This value is a reasonable approximation of what is known about the dust sublimation temperature. 
The temperature and optical properties of dust in a particular radiation field depend on its composition, size, and shape as well as on the density of the surrounding gas \citep[e.g.][]{draine1984,guhathakurta1989}. 
Figure 1 of \citet{baskin2018} depicts how the sublimation temperature of silicate and graphite dust depends on the gas density. For silicate, it varies from 1000 K (at very low gas density) to 2100 K (at extremely high gas density), and for graphite, it is higher by $\sim$300 to 500 K, compared to silicate for the same density range. \citet{baskin2018} argued that for densities most likely met in BLR ($\sim 10^{11}$ cm$^{-3}$), the choice of 2000 K is justified. Since we do not investigate the details of the internal structure of the dusty cloud, our choice of dust sublimation temperature, that is,   2000 K, is a simple and reasonable approximation.

Since \cloudy \ does not automatically check the dust sublimation, we checked the dust temperature provided by the code for each solution. If the dust temperature at the cloud surface was higher than the sublimation temperature, the solution was replaced with a dustless solution. In principle, clouds that are dustless at the irradiated surface can still contain dust in their interior, but a self-consistent solution for such a case from the \cloudy\ code is not possible. Thus, our clouds are either completely dusty or dustless. This is an oversimplification of the task, but it is easy to perform and still allows for some insight into the potential contribution of dust to the cloud heating and cooling as well as radiative transfer under BLR-relevant conditions.

As the output, we stored the fluxes selected at the same bands as in \cite{2019MNRAS.489.5284K}. We also stored the intensities of the selected lines, including H$\beta \, \lambda4861$, Mg~II $\lambda 2795 $, He~II $\lambda 1640$, and Lyman-alpha (Ly$\alpha \, \lambda1215$). In addition, we parameterised and stored the Balmer and Paschen edge, as these features contaminate the time delay measurements in the accretion disc RM \citep[e.g.][]{netzr2022}. Balmer and Paschen edges are the broad spectral features that are understood to represent the difference of intensity of the continuum spectrum on either side of the limit of the Balmer and Paschen series of hydrogen. These edges signify the direct ionisation of the hydrogen atom from the second and third energy levels, correspondingly. Since the series forms an intrinsically broad structure, the depth is measured at some distance from the actual limit. In our measurements, we adopted the wavelength pairs 3620~$\AA$ and 3700~$\AA$ for the Balmer jump, while we adopted 8160~$\AA$ and 8260~$\AA$ for the Paschen jump.

\section{Results}\label{sect:results}
\subsection{Properties of the reprocessing broad-line region continuum}
We computed the ratios of diffuse continuum (DC) emission at four wavelengths (1461~$\AA$, 3641~$\AA$, 5200~$\AA$, and 8100~$\AA$) to the incident continuum emission at 1215~$\AA$, as done in \cite{2019MNRAS.489.5284K}, for both the dustless and dusty solutions. The diffuse continuum emission includes contributions from the reflected incident continuum and the diffuse continuum emission from the outward-facing cloud face. The logarithm contours of these ratios are plotted in the cloud gas density-incident ionising photon flux ($\log n_{\rm H}- \log \Phi_{\rm H}$) plane for dustless (left panel) and dusty BLR (right panel) solutions in Figure \ref{fig:contours}. In the right panel of the figure, the shaded region represents the solutions for the models involving dust while the rest of the part is replaced with dustless solutions. 

Our contours for the dustless solutions are similar to those obtained by \cite{2019MNRAS.489.5284K}. Computations of the dusty clouds appeared to be more complicated, and the solutions did not always converge. This mostly happened for cloud sets that had extremely high dust temperatures in the front layers and low temperatures in the outer layers. Such clouds were not usually self-consistent (dust would have been evaporated in the front layers, thus changing the transmitted spectrum through these zones), and according to our approach, we considered these clouds as `no dust solutions' and replaced them with self-consistent dustless clouds. In this way, we underestimate the role of dust.

We include the lines in our discussion of the continua since some lines do actually form pseudo-continua, for example, Fe II lines \citep{Baldwin_etal_2004, Bruhweiler_Verner_2008, Panda_etal_2018, Panda_etal_2019, Panda_PhD_2021, czerny2023}. We illustrate the issue in Figure~\ref{fig:spectra}. The intense lines, such as  H$\beta$ and Mg II, are clearly seen as separate lines, and the remaining lines contribute to what has traditionally been described as the `small blue bump', which consists mostly of Fe II and Balmer continuum.

\begin{figure}
\includegraphics[height=8cm,width=8cm]{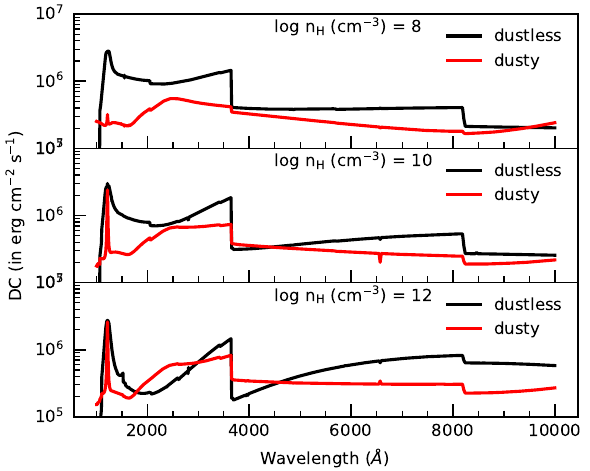}
\caption{\label{fig:spectra_examples}Examples of the continua for dusty and dustless clouds for $\log \Phi_{\rm H} (\rm cm^{-2} s^{-1}) = 17$ and three values of the local density. }
\end{figure}

\begin{figure*}
\centering
\includegraphics[width=8cm, height=8cm]{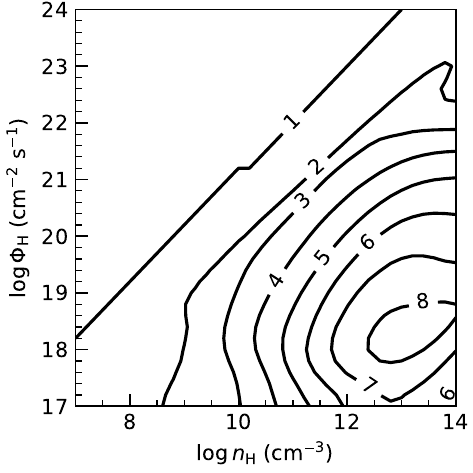}\includegraphics[width=8cm, height=8cm]{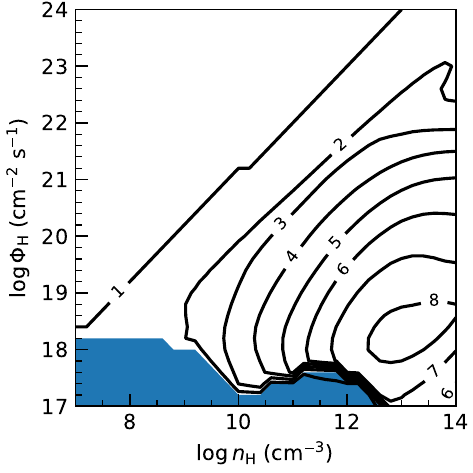}
\caption{\label{fig:drops}Logarithm contours of the ratio, $D$, of the Balmer to Paschen jump (see Equation~\ref{eq:D}) for dustless (left panel) and dusty (right panel) models. The shaded region in the right panel represents the solutions for the dusty cloud for a sublimation temperature of 2000 K. }
\end{figure*}

The shaded region in Figure~\ref{fig:contours} was obtained assuming the sublimation temperature of 2000 K. The region covers a relatively narrow strip of the lowest values of $\log \Phi_{\rm H}$,  mostly below 18.0, but it covers a relatively broad range of densities, from the smallest values adopted in the grid up to $\sim 10^{13}$ cm$^{-3}$. Above this density, dusty solutions do not form because the temperature of the dust grains exceeds the sublimation temperature. 
If a 1500 K value is adopted, the region shrinks further, as the dusty clouds would then be found mostly at fluxes smaller than the minimum value adopted in the grid. The incident flux in the dusty region seems rather small for the BLR modelling, but we address this issue later when discussing the line production in the medium.

The contours of the continuum in the dusty region were modified with respect to dustless solutions. In general, the continuum drops in the presence of the dust. However, at a low density, n$_{\rm H}$ = 10$^8$ cm$^{-3}$, the continuum for the dusty cloud becomes comparable to that of the dustless cloud at around 9500 \AA \ and increases further for longer wavelengths. Also, at a higher density, n$_{\rm H}$ = 10$^{12}$ cm$^{-3}$, the continuum for the dusty cloud enhances in the wavelength range $\sim$1800-3000 \AA. We illustrate this by plotting the continua (see Figure \ref{fig:spectra_examples}) for a few selected models in both dustless and dusty cases while keeping the other parameters constant. The spectral features are strongly affected for the lowest density: the Balmer and Paschen discontinuity almost disappear for a density of$10^8$ cm$^{-3}$ in the presence of the dust. At higher densities, the Balmer edge always increases, while the Paschen edge is almost unaffected. 
\begin{figure*}
\centering
    \includegraphics[width=8cm, height=8cm]{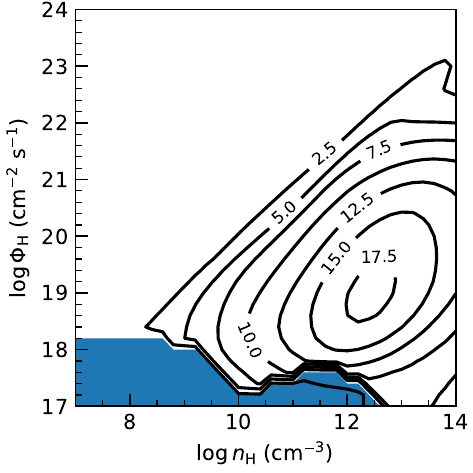}\includegraphics[width=8cm, height=8cm]{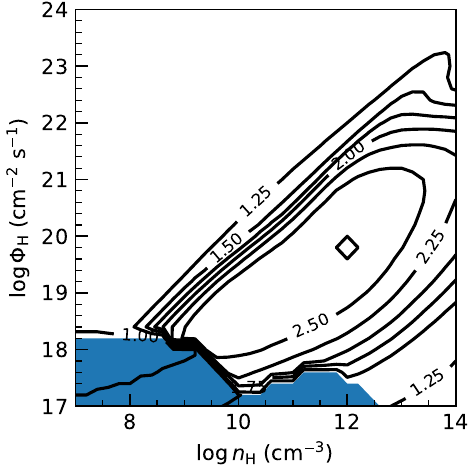}
    \caption{\label{fig:jumps_height}Contours of the Balmer height (left) and Paschen height for dusty (right) models. The blue-shaded region represents the solutions for the dusty cloud for a sublimation temperature of 2000 K.  }
\end{figure*}
We studied the drop of the spectrum at the Balmer and Paschen edge separately. In time delay studies of NGC 5548, the drop in the time delay at the Paschen edge is quite noticeable, and it is possibly deeper than the drop at the Balmer edge \citep{Fausnaugh2016, pei2017}. Therefore, we specifically wanted to investigate the relationship between the Balmer and Paschen edges in the spectrum itself. We parameterised the relative drop, $D$, as the ratio of the Balmer edge to the Paschen edge, that is,
\begin{equation}
\label{eq:D}
D =  \left. \left(\frac{F_{Balmer}^{+}}{F_{Balmer}^{-}}\right) \right/ \left(\frac{F_{Paschen}^{+}}{F_{Paschen}^{-}}\right)
,\end{equation}
where $D$ is the dimensionless quantity. The terms F$_{Balmer}^{+}$ and F$_{Balmer}^{-}$ are the fluxes near the peak (at 3620 $\AA$) and bottom (at 3700$\AA$) of the Balmer jump. Similarly, F$_{Paschen}^{+}$ and F$_{Paschen}^{-}$ are the flux values near the peak (at 8160 $\AA$) and bottom (at 8260 $\AA$) of the Paschen jump.  We plot the corresponding contours of $D$ in Figure~\ref{fig:drops}.

As can be seen from the plots, the ratio $D$ is always larger than one in the dusty parts of the BLR, and in general, it is larger than one for larger densities. A reverse trend (Pashen edge larger than Balmer edge) is only seen for dustless regions, low densities, and extremely high illumination (see Figure~\ref{fig:drops}). We plot the contours for the Balmer and Paschen edges separately, concentrating on the dusty region, in Figure~\ref{fig:jumps_height}. The contour plots for both edges are similar, usually with the Balmer edge being much more prominent. 
\begin{figure*}
\includegraphics[width=9cm, height=9cm]{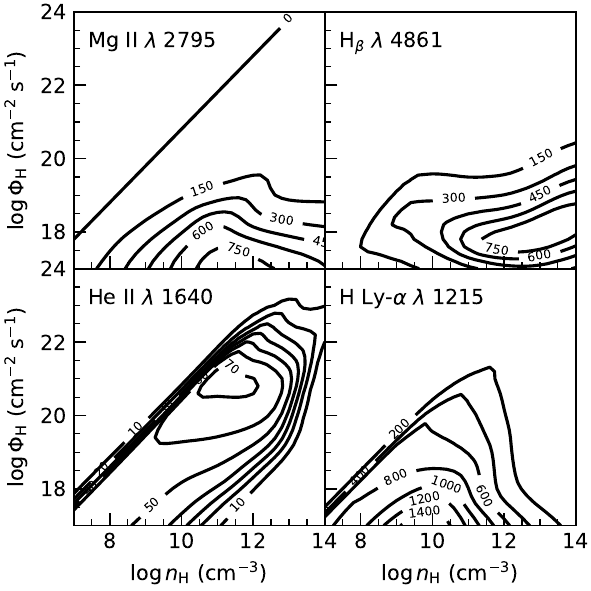}\hfill
\includegraphics[width=9cm, height=9cm]{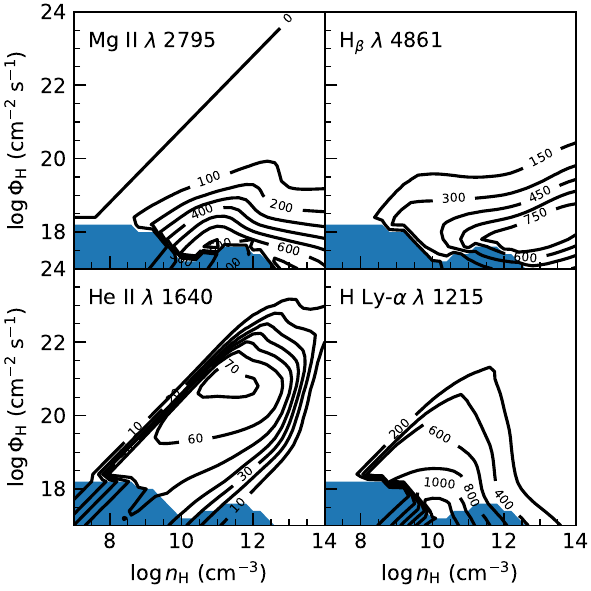}
\caption{\label{fig:dusty_EW_contours}Equivalent widths contours for different emission lines for dustless (left) and dusty (right) solutions. The EWs for Mg~II, He~II, and Ly$\alpha$ are measured with respect to the incident continuum flux at their line centers, while for H$\beta$, EW is measured with respect to the 5100~$\AA$ continuum. }
\end{figure*}
\begin{table}[]
    \centering   \caption{\label{tab:EWS_observed}Observationally determined representative EWs of selected lines in NGC 5548. The UV data are from \citet{1998ApJ...495..718G}, and the H$\beta$ is the average value from \citet{2002ApJ...581..197P}.}
    \begin{tabular}{|c|c|}\hline
      Line   & EW [\AA] \\ \hline%
     Mg~II (2795~\AA)    &  60.9 \\
     H${\beta}$ (4861~\AA) &  79.8 \\
     He~II (1640~\AA) &  9.3 \\
     %C IV (1548 \AA) &  106.5   \\
     Lyman-alpha (1215~\AA) & 114 \\
     \hline
    \end{tabular}      
\end{table}

\subsection{Properties of the emission lines}\label{sect:lines}
The dusty region in Figure~\ref{fig:contours} is present only at very low values of the incident flux, $\Phi_{\rm H}$, which might imply that the BLR with such parameters cannot be responsible for the low ionisation lines (LILs), such as H$\beta$ or Mg II. We therefore calculated the line intensities from our grid of \cloudy \ models with and without dust and determined their equivalent widths (EWs) with respect to the incident continuum. The EWs for Mg II (2795 $\AA$), He II (1640 $\AA$), and Ly$\alpha$ (1215 $\AA$) are measured with respect to the incident continuum flux at their line centers, while for H$\beta$, the EW is measured with respect to the 5100 $\AA$ continuum. Since the models are calculated in plane-parallel geometry, the calculated EWs correspond to the covering factor of 100\%. The corresponding contours are shown in Figure~\ref{fig:dusty_EW_contours}.

We compared the calculated EW values with the observed values of the EWs for NGC 5548, which  we collected from the literature and summarise in Table~\ref{tab:EWS_observed}. The measurements of the UV lines come from the old IUE observations \citep{1998ApJ...495..718G}, and the value of H$\beta$ was calculated as the mean value from the 13-year observational campaign by AGN Watch \citep{2002ApJ...581..197P}. We observed that the EW values for H$\beta$ are comparable to the measured values in NGC 5548 or are much smaller when the incident radiation flux $\log \Phi_{\rm H}$ is larger than 20 cm$^{-2}$ s$^{-1}$. Since the measured EW should include a covering factor $f_c$, which is usually considered to be of the order of 0.1 to 0.3 \citep{Baldwin1995, Korista_Goad_2000, 2019MNRAS.489.5284K, Panda_2021, Panda_2022}, the observed EW can be converted into the model-predicted EW as follows:
\begin{equation}
  EW_{model} = EW_{obs}/f_{c}. 
\end{equation}
By applying the covering factor, the H$\beta$ EW from the model should be on the order of 266 to 798 \AA. Values higher than 750 \AA~ are found in Figure~\ref{fig:dusty_EW_contours} in the dustless region. Thus, if the covering factor $f_c$ is indeed as low as 0.1, the dust present in the BLR is not possible. However, if $f_c$ is higher than 0.1, the dusty solution becomes possible. If $f_c$ is instead on the order of 0.2 to 0.3 (EW requested from the model is then $\sim 300$ to 400 \AA), there is a large parameter region with $ n_{\rm H} \sim 10^{11} - 10^{12}$ cm$^{-3}$ and $\Phi_{\rm H} \sim 3 \times 10^{17}$ cm$^{-2}$ s$^{-1}$ that can give such a line intensity (lower density favoured for higher covering factor). The high density requested for the LIL region is not surprising, as many recent works have obtained such values on the basis of modelling line properties \citep{Bruhweiler_Verner_2008, Panda_etal_2018, Panda_etal_2019, Panda_etal_2020, Panda_2021, Marziani_etal_2021, Sniegowska_etal_2021, Panda_2022, Garnica_etal_2022, Marziani_etal_2023} or just by theoretical argument of radiation pressure confinement \citep{baskin2018}.
Whether the region is dusty or dustless, the values of $\Phi_{\rm H}$ should be smaller than 10$^{19}$ cm$^{-2}$ s$^{-1}$  for the density $10^{12}$ cm$^{-3}$ and smaller than $6 \times 10^{19}$ cm$^{-2}$ s$^{-1}$ for an unlikely high cloud density $10^{14}$ cm$^{-3}$ in order to match the observed EW of H$\beta$ line.

For the Mg II line, the covering factor 0.1 - 0.3 would require the model-predicted EW to be on the order of 200 to 600 \AA. Such values are easily found for dustless clouds when the $\Phi_{\rm H}$ is smaller than 10$^{19}$ cm$^{-2}$ s$^{-1}$ and the density is higher than $10^9$ cm$^{-3}$.
In the case of the dusty region, such conditions are met only if the irradiation is very low ($\Phi_{\rm H} < 10^{17.5}$ cm$^{-2}$ s$^{-1}$). The required density is high (above $\sim 10^{10}$ cm$^{-3}$), and the values of the line EW hardly reach 600 \AA~ in the dusty solutions. So again, as in the case of the H$\beta$ line, dusty solutions require $f_c$ to be above 0.1.

Since the MgII and H$\beta$ both belong to the LIL, they could be emitted by roughly the same region, which would imply the same value of the ionisation flux, density, and covering factor, reproducing the requested EWs for both H$\beta$ and Mg II. We thus searched our grid of solutions for pairs that satisfied these requirements in the best way using an $\chi^2$-selection method. We found two minima. The formally better solution requires the following parameters: $f_c = 0.25$, $\log \Phi_{\rm H}$ (cm$^{-2}$ s$^{-1}$) = 18.8, $\log$ n$_{\rm H}$ (cm$^{-3}) = 10$. This region corresponds to a dustless BLR. However, we also found a second (local) minimum that is in the dusty region. It requires a somewhat lower covering factor, 0.14; lower irradiation ($\log \Phi_{\rm H}$ (cm$^{-2}$ s$^{-1}$) = 17.6); and a higher density ($\log$ n$_{\rm H}$ (cm$^{-3}) \sim 11.4$). So, basically, the dustless BLR is favoured, as the primary minimum is significantly deeper. We discuss the issue further in the next section.

For the high ionisation line (HIL) region, we can find a solution where He II and Ly$\alpha$ come from the same plasma and the region is closer to the black hole. It corresponds to $\log \Phi_{\rm H}$ (cm$^{-2}$ s$^{-1}$) $\sim$ 19, $\log$ n$_{\rm H}$ (cm$^{-3}$) $\sim 10$, and the region is dustless. The corresponding covering factor is of the same order, $f_c \sim 0.2.$

\section{Combining the reprocessing results with the time delay constraints for the effective radius of the high ionisation and low ionisation lines}\label{sect:with_delay}
The time delay for the $H\beta$ line in NGC 5548 has been extensively studied over the years. 
In the case of the H$\beta$ line, for consistency, we used the mean time delay from the AGN Watch campaign \citep{2002ApJ...581..197P}, which is $17.0 \pm 3.9$ days. We used the centroid time delay, and we calculated the weighted mean and the corresponding dispersion by symmetrisation of the errors.

If the BLR receives the same radiation as the observer, then the incident photon flux at this distance should be
\begin{equation}
\Phi = {1 \over 2 \pi R_{BLR}^2}~~\int^{\infty}_{13.6 eV} {L_{\nu} \over h \nu} d\nu, 
\end{equation}
which for the incident radiation flux normalised to the ionising luminosity of $1.17 \times 10^{44}$ erg s$^{-1}$, as estimated by \citet{mehdipour2016} for unabsorbed flux luminosity, and the distance corresponding to H$\beta$ time delay is equal to $(4.02\pm2.11) \times 10^{17}$ photons cm$^{-2}$ s$^{-1}$, which is about an order of magnitude lower than the value of the incident flux ($\Phi_{\rm H} =6.3 \times 10^{18}$ cm$^{-2}$ s$^{-1}$) required to reproduce both the H$\beta$ and the Mg II lines in the dustless case (see Section~\ref{sect:lines}). This low value would eventually locate the emitting region within the dusty part, depending on the local density. It is actually consistent with the secondary minimum we found in the dusty region and strongly in contradiction with the primary minimum, implying the dustless solution.

Therefore, if we assume that the H$\beta$ producing clouds has the same incident continuum, which is not blocked by any obscurer, and the distance corresponds to the measured time delay, then this region is dusty. The favoured density is high, $\log n_{\rm H}$ (cm$^{-3}$) $\sim 11.4$, and a relatively small covering factor of 0.14 is enough to reproduce the EW of the observed line. 
\citet{gaskell2007} found a covering factor of 40\%, but the analysis included the need for large intrinsic reddening of the source by $E(B-V) = 0.17$ mag and a typical extinction curve different from an SMC type. A lower covering factor could be achieved by taking higher densities and including the high internal turbulent velocity \citep[see e.g.][]{Panda_2021}. The extinction advocated by \citet{gasell2023} underestimates the ionising flux by a factor of approximately seven.

The time delay for Mg II is not known so well. The early IUE measurement by \citet{clavel1991} gave the delay as being on the order of 34 to 72 days, and \citet{cackett2015} were not able to measure the Mg II delay from their campaign. We can thus use the general Radius-Luminosity relation for the Mg II recently obtained from other sources \citep{czerny_CTS_2019,zajacek2020,zajacek2021}. The general relation from \citet{zajacek2021} reads
\begin{equation}
\log \tau = 0.3 \log L_{3000} + 1.67,
\end{equation}
where $\tau$ is the time delay measured in days and $L_{3000}$ is the luminosity at 3000 $\AA$ in the units of $10^{44}$ erg s$^{-1}$. Adopting the logarithm of the luminosity at 3000 \AA~ of 43.67 erg s$^{-1}$ from \citet{dovciak2022}, we obtained a time delay of 37 days, which is longer than for H$\beta$ by a factor of two. Adopting the luminosity from \citet{lawther2018} of 43.58 erg s$^{-1}$ and the radius-luminosity relation from \citet{Yu2023}, we obtained a time delay of 33 days, which is only marginally shorter. So there is indeed some stratification even within the LIL part. 

We thus searched for the best representation of the EWs of the H$\beta$ and MgII lines, assuming that the Mg II site should be two times farther than the H$\beta$ site. We found a pair of parameters that best matches the observed EWs at the density of $\log n_{\rm H}$ (cm$^{-3}$) = 12 and the ionisation flux $\log \Phi_{\rm H} (\rm cm^{-2} s^{-1})$ of 19.2 for H$\beta$ and 18.60 for Mg II, respectively. This solution, however, is not consistent with the measured time delays. The second minimum was found at  $\log n_{\rm H}$ (cm$^{-3}$) = 10 for a pair of ionisation  fluxes $\log \Phi_{\rm H} (\rm cm^{-2} s^{-1})$ of 17.60 for H$\beta$ and 17 for Mg II, respectively.  This solution is fully consistent with the measured H$\beta$ time delay, as discussed above. The density is relatively low (the density was estimated as $\log n_{\rm H}$ (cm$^{-3}$)  = 11.4 when we requested the Mg II and H$\beta$ regions to have the same distance from the black hole), so it locates the solution right at the edge of the dusty zone. This may imply that the clouds are actually partially dusty, but such an option cannot be included in our model yet. The drop in the density was forced by the rise in the covering factor, which was equal to 0.24 for this pair of solutions. The contours of the EWs of the H$\beta$ and Mg II lines are not identical, so combining pairs at different locations (i.e. incident flux) requires a corresponding change of the density if the density and the covering factor are to remain the same for both lines. In general, this does not have to be the case, but without such a requirement, we cannot constrain the solutions.

The effective location of the Mg II and H$\beta$ is not, however, set very firmly. The delay of $H\beta$ is directly measured as $17.0 \pm 3.9$ days (the mean value, as discussed above), but the actual delay changes from year to year. The delay of Mg II was not measured in NGC 5548, and the value of 37 days comes from the R-L relation with a dispersion of 0.30 dex \citep{zajacek2021}, so it should roughly read as $37^{+31}_{-20}$ days, and the two regions may well be the same within the error or separated by a factor four. Obtaining more stringent limits would require direct measurement of the Mg II line delay in this source.

\section{The role of obscuration of the incident radiation flux for the broad-line region properties and the dusty-dustless transition in the broad-line region}\label{sect:filtering}
As discussed in several papers, the lines do not receive the same incident continuum as we do \citep{holidays2019, Panda_2021, dovciak2022}. The most discussed recent phenomenon observed in NGC 5548 is the `line holidays' seen in HILs such as C IV, Si IV, and He II \citep{goad2016,holidays2019} but also in H$\beta$ \citep{pei2017}. Usually, the BLR emission line variations are well correlated with changes in the incident ionising continuum with a time lag, but during `line holidays' such correlations are missing. To explain this phenomenon, an obscurer, located between the nucleus and the BLR, is proposed as affecting the observed continuum \citep{mehdipour2016,kriss2019,deghanian2019,wildy2021}.

 In their simulations, \citet{holidays2019} considered the role of the equatorial obscurer, which was assumed to have $N_H = 10^{23}$ cm$^{-2}$, $\Phi_{\rm H} = 10^{20.3}$ cm$^{-2}$ s$^{-1}$ (but the distance is unknown), and the local densities $10^9 - 10^{12}$ cm$^{-3}$. We followed their assumptions about the ionisation parameter and the column density, selected the value of the density as $2 \times 10^{10}$ cm$^{-3}$, obtained the transmitted spectrum, and used it as the incident flux of the BLR with the aim of examining if the dust is more likely to be present if the continuum is attenuated by an absorber located between the central parts of the disc and the BLR.

\begin{figure}
\includegraphics[width=8cm, height=8cm]{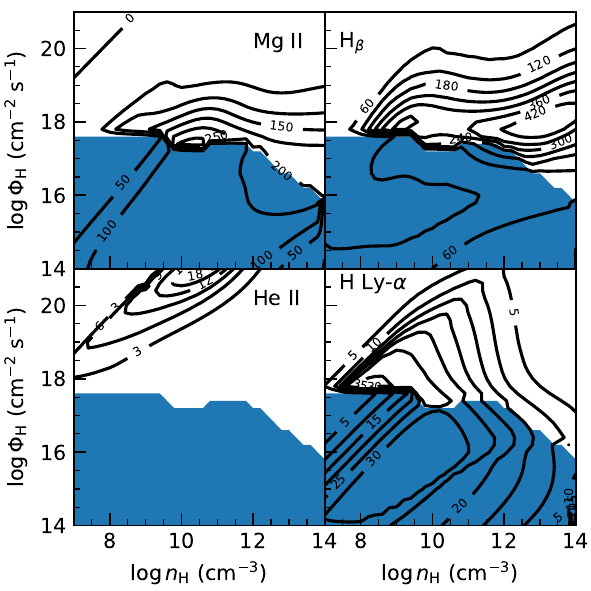}
\caption{\label{fig:obscured_EW_contours}Equivalent width contours for different emission lines for dusty solutions when considering an obscurer with $\log \Phi_{\rm H} (\rm cm^{-2} s^{-1}) = 20.3, \log$ n$_{\rm H}$ (cm$^{-3}$) = 10.3.}

\end{figure}

The new contour plots for emission lines are shown in Figure~\ref{fig:obscured_EW_contours}. In the figure, we used the shifted range of the incident flux since then the lower values might be of more interest. The dusty part is not very different since the ionising flux measures the normalisation of the incident radiation onto the cloud. The contours are slightly affected by the intervening absorber since the shape of the incident radiation has changed in comparison with that used in Figure~\ref{fig:dusty_EW_contours}. 

We compared the results for H$\beta$ with the data corresponding to the line holidays, although the H$\beta$ line did not seem strongly affected in the study by \citet{deghanian2019}. We estimated the mean EW of the H$\beta$ line, EW$_{H\beta}$= 63.35$\pm$0.04 $\AA$, taking data from \cite{pei2017} for the period from  2014 January 5 to 2014 August 5 including the BLR holiday. The ionising flux at a given location dropped by a factor of 3.27 in comparison with the previous case. However, during this period the H$\beta$ emission line light curve lags in relation to the continuum light curve at 5100 $\AA$ by 4.17$\pm$0.36 days \citep{pei2017}, so the effective distance of the cloud was reduced by a factor  4.08, and the incident radiation flux effectively increased by a factor 5.08. Subsequently adopting $\log \Phi_{\rm H}$ (cm$^{-2}$ s$^{-1}$) = 18.3 as describing the H$\beta$ region, we observed that this line must then come from dustless clouds. Values in the broad range of EW are available, and they strongly depend on the density, up to 420 \AA, so the required value of 63.35 can be easily accommodated, as the covering factor is as low as $\sim 15$\%.

\section{Discussion} \label{sect:discussion}
The possibility of dust presence in the BLR is an open issue, and we addressed it by performing the calculation of the radiation reprocessing by the BLR material in two cases, that is, with and without the presence of dust. The dust affects the shape of the reprocessing continuum, the depth of the Balmer and Paschen edges, and the line emissivity. 

The emission lines were not as strongly suppressed as suggested by \citet{Netzer_Laor_1993} when we assumed the large local density of the BLR clouds (above $\sim 10^9$ cm$^{-3}$). A similar conclusion was reached by \citet{adhikari2016} for the intermediate line region. We performed a comparison of the EWs for the Mg II and H$\beta$ lines from the model with the data for NGC 5548, taking into account the observed ionising flux calculated from the observed SED and the distance of these lines from the black hole. The dusty solution consistently represents the data, and the covering factor is 0.14. We did not see any budget problem \citep[see][]{gaskell2007}, and to represent the mean values of Mg II and H$\beta$, no obscurer is needed. The HIL region responsible for HILs (for example, the He II line) is located much closer to the central source and thus must be highly ionised and transparent enough to not affect the H$\beta$ and Mg II regions.

Our results therefore support the BLR models of \citet{czhr2011} and \citet{baskin2018}, which require a certain amount of dust to be present in order to alleviate the accretion disc material and provide the gas for the BLR. This dust is important for the BLR dynamics since it launches (mostly) a failed wind in the first model and puffs up the disc, allowing for efficient irradiation in the second model. This is an important conclusion since, in the case of a dusty BLR, the inner radius of the low ionisation part of the BLR is universally set by the dust sublimation temperature, and attempts to use the radius-luminosity relation for cosmology are justified \citep{MartinezAldama2019,2019FrASS...6...75P,zajacek2020,czerny_acta2021,zajacek2021,khadka2021,cao2022,czerny2023}.

The obscurer seen in NGC 5548 during the line holidays must then indeed be temporary since the mean line intensity is well reproduced self-consistently with their location. However, the obscurer period reveals an interesting subject of dusty-dustless transitions in the region where the H$\beta$ line originates. When the H$\beta$ region is strongly moved inwards during the obscuration period, dust cannot form since the dust presence in the accretion disc atmosphere is mostly set by the local dissipation, which cannot change in a timescale of months \citep[see e.g.][for a short review]{czerny_review2006}. In this case, the emission likely comes from the irradiated disc atmosphere at the required radius. It might seem surprising that the covering factor did not change with the change in the distance and formation mechanism of the BLR. However, as shown by \citet{naddaf2021,naddaf2022}, when the source is characterised by the low value of the Eddington ratio (as is the case for NGC 5548), only the dust-driven failed wind forms, with small cloud vertical velocities, so the clouds then actually simply form a turbulent surface right above the accretion disc.

\subsection{No locally optimised cloud in our computations}
In general, the geometry and the content of the BLR may be complex, with a range of radii involved as well as a range of densities at each radius. To represent this complexity, the locally optimised cloud (LOC) approach is recommended \citep{Baldwin1995, Bottorf2002, DuShou2023}. However, in our comparison with the NGC 5548 data, we include the radial stratification of the BLR known from RM studies. Thus, we do not account for a range of local densities at a given distance, as is done in the LOC approach. However, arguments based on radiation pressure confinement \citep{baskin2018} and/or thermal instability in the irradiated medium \citep{krolik1981,begelman1990,agata1999,agata2006,agata2017} rather favour a specific density on the order of $10^{11}$ cm$^{-3}$ when the medium becomes clumpy.

\subsection{Dusty cloud stratification}
In our approach, we only considered  clouds as being either dustless or dusty across the cloud. We did not include solutions produced by \cloudy \ where the dust was formally contained but the cloud surface was hotter than the adopted sublimation temperature, even if the temperature ultimately dropped inside the cloud with the attenuation of radiation. We could not consider clouds that are partially dusty, that is, dustless in the illuminated face and dusty in the interior and the dark face. \cloudy \ does not allow for such solutions, and the solutions with irradiated faces hotter than the sublimation temperature do not transfer the radiation to the cloud interior in the correct way. Division of the cloud into dusty and dustless is not simple since the returning radiation is then not treated properly. Complex iterations would be necessary, and they are beyond the scope of this current pilot paper.

\subsection{Variability and obscuration}
The source NGC 5548 is known to have been highly variable over the years with respect to both the continuum as well as emission lines \citep[e.g.][]{peterson1987,clavel1991,peterson1991,rokaki1993,shapovalova2004, Sergeev_etal_2007,edelson2015, DeRosa_etal_2015,mehdipour2016, pei2017, Bon_etal_2018, Horne_etal_2021, Panda_etal_2022}. The study by \citet{mao2018} of {\it XMM-Newton}'s Reflection Grating Spectrometer (RGS) data from the years of 2013 to 2014 and 2016 did not show any variability, but it was only sensitive to the highly ionised absorber located along the line of sight at a distance of 1 pc or more from the black hole.

Interesting aspects of variability were noticed by \citet{chiang2000}. They showed that the variability is led by EUV (at $\sim 0.2 $ kev, with harder X-rays (above $\sim 1 $ keV) lagging by 10 to 30 ksec. 

Our study addresses only the mean BLR properties of NGC 5548 and, separately, a period of line holidays. The source variability, however, implies that the BLR cannot be well treated within the frame of a stationary model, particularly in the dusty part. Dust forms only at the disc surface \citep{elvis2002,czhr2011}, where the pressure is high enough, and the disc changes mostly in the viscous timescale, with some effects of the irradiation, which reflects much faster variability at the disc inner radius combined with the light travel time. In clouds, dust cannot form since the gas temperature in the clouds is of the order of $10^4$ K. The dust grains embedded in the clouds can exist there if their temperature is below the sublimation temperature since the dust cools roughly as a black body, while the gas cooling by lines is much less efficient. Dust grains in the clouds can be easily destroyed if the irradiating flux rises, but they cannot reform if the irradiation drops. However, such a hysteresis effect will be quite complex to reproduce in the model.

\section{Conclusions}\label{sect:conclusion}
In this work, we investigated how the dust present in the BLR influences the reprocessed radiation by using a set of \cloudy \ models for dusty and dustless BLR clouds. We tested our model for NGC 5548 by comparing the model-predicted EWs with the observed EWs for a set of BLR emission lines. The key points of our findings are outlined below:
\begin{enumerate}
    \item  Dust may exist in the BLR for a narrow range of the incident ionising flux ($\Phi_{\rm H} < 10^{18}$cm$^{-2}$ s$^{-1}$) and a comparatively large range of local density (n$_{\rm H}$ up to 10$^{13}$ cm$^{-3}$). 
    \item In the dusty region of the BLR, the Balmer and Paschen discontinuities almost completely vanish at low density ($\sim$ 10$^8$ cm$^{-3}$). At higher densities, the Paschen edge is essentially unaffected, but the Balmer edge constantly grows.
    \item In the presence of dust, the Balmer edge is always larger than the Paschen edge.
    \item When combining our reprocessing results with time-delay measurements, we observed that the LILs originate in the dusty BLR region, and there is no need for shielding between the inner disc and BLR.
    \item When there is an extra absorber present, the H$\beta$ emitting region shifts towards the part of the accretion disc surface that is free of dust.
\end{enumerate}
\begin{acknowledgements}
We gratefully acknowledge the useful feedback provided by the anonymous referee, whose recommendations and comments improved the work. This project has received funding from the European Research Council (ERC) under the European Union’s Horizon 2020 research and innovation program (grant agreement No. [951549]). Part of this work was supported by the Polish Funding Agency National Science Centre, project 2017/26/A/ST9/00756 (MAESTRO 9), and by OPUS-LAP/GA CR-LA bilateral project
(2021/43/I/ST9/01352/OPUS 22 and GF23-04053L). SP acknowledges the Conselho Nacional de Desenvolvimento Científico e Tecnológico (CNPq) Fellowships 164753/2020-6 and 300936/2023-0. MZ acknowledges the financial support by the GACR EXPRO grant no. GX21-13491X. M.L.M.-A. acknowledges financial support from Millenium Nucleus NCN19-058 (TITANs).
\end{acknowledgements}

\bibliographystyle{aa} % style aa.bst
\bibliography{master} % your references 

\begin{thebibliography}{132}
\expandafter\ifx\csname natexlab\endcsname\relax\def\natexlab#1{#1}\fi

\bibitem[{{Adhikari} {et~al.}(2016){Adhikari}, {R{\'o}{\.z}a{\'n}ska},
  {Czerny}, {Hryniewicz}, \& {Ferland}}]{adhikari2016}
{Adhikari}, T.~P., {R{\'o}{\.z}a{\'n}ska}, A., {Czerny}, B., {Hryniewicz}, K.,
  \& {Ferland}, G.~J. 2016, \apj, 831, 68

\bibitem[{{Alonso-Herrero} {et~al.}(2012){Alonso-Herrero}, {Pereira-Santaella},
  {Rieke}, \& {Rigopoulou}}]{alonso_herrero_2012}
{Alonso-Herrero}, A., {Pereira-Santaella}, M., {Rieke}, G.~H., \& {Rigopoulou},
  D. 2012, \apj, 744, 2

\bibitem[{{Alonso-Herrero} {et~al.}(2003){Alonso-Herrero}, {Quillen}, {Rieke},
  {Ivanov}, \& {Efstathiou}}]{alonso_herrero2003}
{Alonso-Herrero}, A., {Quillen}, A.~C., {Rieke}, G.~H., {Ivanov}, V.~D., \&
  {Efstathiou}, A. 2003, \aj, 126, 81

\bibitem[{{Antonucci} \& {Miller}(1985)}]{antonucci1985}
{Antonucci}, R.~R.~J. \& {Miller}, J.~S. 1985, \apj, 297, 621

\bibitem[{{Asmus} {et~al.}(2016){Asmus}, {H{\"o}nig}, \&
  {Gandhi}}]{Asmus_etal_2016}
{Asmus}, D., {H{\"o}nig}, S.~F., \& {Gandhi}, P. 2016, \apj, 822, 109

\bibitem[{{Baldwin} {et~al.}(1995){Baldwin}, {Ferland}, {Korista}, \&
  {Verner}}]{Baldwin1995}
{Baldwin}, J., {Ferland}, G., {Korista}, K., \& {Verner}, D. 1995, \apjl, 455,
  L119

\bibitem[{{Baldwin} {et~al.}(2004){Baldwin}, {Ferland}, {Korista}, {Hamann}, \&
  {LaCluyz{\'e}}}]{Baldwin_etal_2004}
{Baldwin}, J.~A., {Ferland}, G.~J., {Korista}, K.~T., {Hamann}, F., \&
  {LaCluyz{\'e}}, A. 2004, \apj, 615, 610

\bibitem[{{Baldwin} {et~al.}(1991){Baldwin}, {Ferland}, {Martin}, {Corbin},
  {Cota}, {Peterson}, \& {Slettebak}}]{1991ApJ...374..580B}
{Baldwin}, J.~A., {Ferland}, G.~J., {Martin}, P.~G., {et~al.} 1991, \apj, 374,
  580

\bibitem[{{Barvainis}(1987)}]{barvainis1987}
{Barvainis}, R. 1987, \apj, 320, 537

\bibitem[{{Baskin} \& {Laor}(2018)}]{baskin2018}
{Baskin}, A. \& {Laor}, A. 2018, \mnras, 474, 1970

\bibitem[{{Begelman} \& {McKee}(1990)}]{begelman1990}
{Begelman}, M.~C. \& {McKee}, C.~F. 1990, \apj, 358, 375

\bibitem[{{Bon} {et~al.}(2018){Bon}, {Bon}, \& {Marziani}}]{Bon_etal_2018}
{Bon}, N., {Bon}, E., \& {Marziani}, P. 2018, Frontiers in Astronomy and Space
  Sciences, 5, 3

\bibitem[{{Bottorff} {et~al.}(2002){Bottorff}, {Baldwin}, {Ferland},
  {Ferguson}, \& {Korista}}]{Bottorf2002}
{Bottorff}, M.~C., {Baldwin}, J.~A., {Ferland}, G.~J., {Ferguson}, J.~W., \&
  {Korista}, K.~T. 2002, \apj, 581, 932

\bibitem[{{Brocklehurst}(1971)}]{broclehurst1971}
{Brocklehurst}, M. 1971, \mnras, 153, 471

\bibitem[{{Bruhweiler} \& {Verner}(2008)}]{Bruhweiler_Verner_2008}
{Bruhweiler}, F. \& {Verner}, E. 2008, \apj, 675, 83

\bibitem[{{Burtscher} {et~al.}(2013){Burtscher}, {Meisenheimer}, {Tristram},
  {Jaffe}, {H{\"o}nig}, {Davies}, {Kishimoto}, {Pott}, {R{\"o}ttgering},
  {Schartmann}, {Weigelt}, \& {Wolf}}]{Burtscher_etal_2013}
{Burtscher}, L., {Meisenheimer}, K., {Tristram}, K.~R.~W., {et~al.} 2013, \aap,
  558, A149

\bibitem[{{Cackett} {et~al.}(2015){Cackett}, {G{\"u}ltekin}, {Bentz},
  {Fausnaugh}, {Peterson}, {Troyer}, \& {Vestergaard}}]{cackett2015}
{Cackett}, E.~M., {G{\"u}ltekin}, K., {Bentz}, M.~C., {et~al.} 2015, \apj, 810,
  86

\bibitem[{{Cao} {et~al.}(2022){Cao}, {Zaja{\v{c}}ek}, {Panda},
  {Mart{\'\i}nez-Aldama}, {Czerny}, \& {Ratra}}]{cao2022}
{Cao}, S., {Zaja{\v{c}}ek}, M., {Panda}, S., {et~al.} 2022, \mnras, 516, 1721

\bibitem[{{Cerqueira-Campos} {et~al.}(2023){Cerqueira-Campos},
  {Rodr{\'\i}guez-Ardila}, {Panda}, {Riffel}, {Dahmer-Hahn}, \&
  {Marinello}}]{Cerqueira-Campos_etal_2023}
{Cerqueira-Campos}, F.~C., {Rodr{\'\i}guez-Ardila}, A., {Panda}, S., {et~al.}
  2023, \mnras, 524, 542

\bibitem[{{Chen} {et~al.}(2023){Chen}, {Liu}, {Zhai}, {Yao}, {Li}, {Du}, {Hu},
  {Guo}, {Xiao}, {Songsheng}, \& {Wang}}]{chen2023}
{Chen}, Y.-J., {Liu}, J.-R., {Zhai}, S., {et~al.} 2023, \mnras, 522, 3439

\bibitem[{{Chiang} {et~al.}(2000){Chiang}, {Reynolds}, {Blaes}, {Nowak},
  {Murray}, {Madejski}, {Marshall}, \& {Magdziarz}}]{chiang2000}
{Chiang}, J., {Reynolds}, C.~S., {Blaes}, O.~M., {et~al.} 2000, \apj, 528, 292

\bibitem[{{Clavel} {et~al.}(1991){Clavel}, {Reichert}, {Alloin}, {Crenshaw},
  {Kriss}, {Krolik}, {Malkan}, {Netzer}, {Peterson}, {Wamsteker}, {Altamore},
  {Baribaud}, {Barr}, {Beck}, {Binette}, {Bromage}, {Brosch}, {Diaz},
  {Filippenko}, {Fricke}, {Gaskell}, {Giommi}, {Glass}, {Gondhalekar},
  {Hackney}, {Halpern}, {Hutter}, {Joersaeter}, {Kinney}, {Kollatschny},
  {Koratkar}, {Korista}, {Laor}, {Lasota}, {Leibowitz}, {Maoz}, {Martin},
  {Mazeh}, {Meurs}, {Nair}, {O'Brien}, {Pelat}, {Perez}, {Perola}, {Ptak},
  {Rodriguez-Pascual}, {Rosenblatt}, {Sadun}, {Santos-Lleo}, {Shaw}, {Smith},
  {Stirpe}, {Stoner}, {Sun}, {Ulrich}, {van Groningen}, \&
  {Zheng}}]{clavel1991}
{Clavel}, J., {Reichert}, G.~A., {Alloin}, D., {et~al.} 1991, \apj, 366, 64

\bibitem[{{Clavel} {et~al.}(1989){Clavel}, {Wamsteker}, \&
  {Glass}}]{clavel1989}
{Clavel}, J., {Wamsteker}, W., \& {Glass}, I.~S. 1989, \apj, 337, 236

\bibitem[{{Collin-Souffrin} {et~al.}(1988){Collin-Souffrin}, {Dyson},
  {McDowell}, \& {Perry}}]{suzy1988}
{Collin-Souffrin}, S., {Dyson}, J.~E., {McDowell}, J.~C., \& {Perry}, J.~J.
  1988, \mnras, 232, 539

\bibitem[{{Czerny}(2006)}]{czerny_review2006}
{Czerny}, B. 2006, in Astronomical Society of the Pacific Conference Series,
  Vol. 360, AGN Variability from X-Rays to Radio Waves, ed. C.~M. {Gaskell},
  I.~M. {McHardy}, B.~M. {Peterson}, \& S.~G. {Sergeev}, 265

\bibitem[{{Czerny} \& {Hryniewicz}(2011)}]{czhr2011}
{Czerny}, B. \& {Hryniewicz}, K. 2011, \aap, 525, L8

\bibitem[{{Czerny} {et~al.}(2017){Czerny}, {Li}, {Hryniewicz}, {Panda},
  {Wildy}, {Sniegowska}, {Wang}, {Sredzinska}, \& {Karas}}]{czerny2017}
{Czerny}, B., {Li}, Y.-R., {Hryniewicz}, K., {et~al.} 2017, \apj, 846, 154

\bibitem[{{Czerny} {et~al.}(2021){Czerny}, {Mart{\'\i}nez-Aldama},
  {Wojtkowska}, {Zaja{\v{c}}ek}, {Marziani}, {Dultzin}, {Naddaf}, {Panda},
  {Prince}, {Przyluski}, {Ralowski}, \& {{\'S}niegowska}}]{czerny_acta2021}
{Czerny}, B., {Mart{\'\i}nez-Aldama}, M.~L., {Wojtkowska}, G., {et~al.} 2021,
  Acta Physica Polonica A, 139, 389

\bibitem[{{Czerny} {et~al.}(2015){Czerny}, {Modzelewska}, {Petrogalli}, {Pych},
  {Adhikari}, {{\.Z}ycki}, {Hryniewicz}, {Krupa}, {{\'S}wie{\c{t}}o{\'n}}, \&
  {Niko{\l}ajuk}}]{czerny2015}
{Czerny}, B., {Modzelewska}, J., {Petrogalli}, F., {et~al.} 2015, Advances in
  Space Research, 55, 1806

\bibitem[{{Czerny} {et~al.}(2019){Czerny}, {Olejak}, {Ra{\l}owski},
  {Koz{\l}owski}, {Martinez Aldama}, {Zajacek}, {Pych}, {Hryniewicz},
  {Pietrzy{\'n}ski}, {Sobrino Figaredo}, {Haas}, {{\'S}redzi{\'n}ska}, {Krupa},
  {Kurcz}, {Udalski}, {Gorski}, {Karas}, {Panda}, {Sniegowska}, {Naddaf},
  {Bilicki}, \& {Sarna}}]{czerny_CTS_2019}
{Czerny}, B., {Olejak}, A., {Ra{\l}owski}, M., {et~al.} 2019, \apj, 880, 46

\bibitem[{{Czerny} {et~al.}(2023{\natexlab{a}}){Czerny}, {Panda}, {Prince},
  {Jaiswal}, {Zajacek}, {Martinez Aldama}, {Kozlowski}, {Kovacevic}, {Ilic},
  {Popovic}, {Pozo Nunez}, {Hoenig}, \& {Brandt}}]{Czerny_etal_2023_LSST}
{Czerny}, B., {Panda}, S., {Prince}, R., {et~al.} 2023{\natexlab{a}}, arXiv
  e-prints, arXiv:2301.08975

\bibitem[{{Czerny} {et~al.}(2023{\natexlab{b}}){Czerny}, {Panda}, {Prince},
  {Kumar Jaiswal}, {Zaja{\v{c}}ek}, {Martinez Aldama}, {Koz{\l}owski},
  {Kovacevic}, {Ilic}, {Popovi{\'c}}, {Pozo Nu{\~n}ez}, {H{\"o}nig}, \&
  {Brandt}}]{czerny2023}
{Czerny}, B., {Panda}, S., {Prince}, R., {et~al.} 2023{\natexlab{b}}, \aap,
  675, A163

\bibitem[{{De Rosa} {et~al.}(2015){De Rosa}, {Peterson}, {Ely}, {Kriss},
  {Crenshaw}, {Horne}, {Korista}, {Netzer}, {Pogge}, {Ar{\'e}valo}, {Barth},
  {Bentz}, {Brandt}, {Breeveld}, {Brewer}, {Dalla Bont{\`a}}, {De
  Lorenzo-C{\'a}ceres}, {Denney}, {Dietrich}, {Edelson}, {Evans}, {Fausnaugh},
  {Gehrels}, {Gelbord}, {Goad}, {Grier}, {Grupe}, {Hall}, {Kaastra}, {Kelly},
  {Kennea}, {Kochanek}, {Lira}, {Mathur}, {McHardy}, {Nousek}, {Pancoast},
  {Papadakis}, {Pei}, {Schimoia}, {Siegel}, {Starkey}, {Treu}, {Uttley},
  {Vaughan}, {Vestergaard}, {Villforth}, {Yan}, {Young}, \&
  {Zu}}]{DeRosa_etal_2015}
{De Rosa}, G., {Peterson}, B.~M., {Ely}, J., {et~al.} 2015, \apj, 806, 128

\bibitem[{{Dehghanian} {et~al.}(2019{\natexlab{a}}){Dehghanian}, {Ferland},
  {Kriss}, {Peterson}, {Mathur}, {Mehdipour}, {Guzm{\'a}n}, {Chatzikos}, {van
  Hoof}, {Williams}, {Arav}, {Barth}, {Bentz}, {Bisogni}, {Brandt}, {Crenshaw},
  {Dalla Bont{\`a}}, {De Rosa}, {Fausnaugh}, {Gelbord}, {Goad}, {Gupta},
  {Horne}, {Kaastra}, {Knigge}, {Korista}, {McHardy}, {Pogge}, {Starkey}, \&
  {Vestergaard}}]{holidays2019}
{Dehghanian}, M., {Ferland}, G.~J., {Kriss}, G.~A., {et~al.}
  2019{\natexlab{a}}, \apj, 877, 119

\bibitem[{{Dehghanian} {et~al.}(2019{\natexlab{b}}){Dehghanian}, {Ferland},
  {Peterson}, {Kriss}, {Korista}, {Chatzikos}, {Guzm{\'a}n}, {Arav}, {De Rosa},
  {Goad}, {Mehdipour}, \& {van Hoof}}]{deghanian2019}
{Dehghanian}, M., {Ferland}, G.~J., {Peterson}, B.~M., {et~al.}
  2019{\natexlab{b}}, \apjl, 882, L30

\bibitem[{{Donnan} {et~al.}(2023){Donnan}, {Rigopoulou}, {Garc{\'\i}a-Bernete},
  {Pereira-Santaella}, {Alonso-Herrero}, {Roche}, {Aalto},
  {Hern{\'a}n-Caballero}, \& {Spoon}}]{donnan2023}
{Donnan}, F.~R., {Rigopoulou}, D., {Garc{\'\i}a-Bernete}, I., {et~al.} 2023,
  \aap, 669, A87

\bibitem[{{Dov{\v{c}}iak} {et~al.}(2022){Dov{\v{c}}iak}, {Papadakis},
  {Kammoun}, \& {Zhang}}]{dovciak2022}
{Dov{\v{c}}iak}, M., {Papadakis}, I.~E., {Kammoun}, E.~S., \& {Zhang}, W. 2022,
  \aap, 661, A135

\bibitem[{{Draine} \& {Lee}(1984)}]{draine1984}
{Draine}, B.~T. \& {Lee}, H.~M. 1984, \apj, 285, 89

\bibitem[{{Du} {et~al.}(2023){Du}, {Zhai}, \& {Wang}}]{DuShou2023}
{Du}, P., {Zhai}, S., \& {Wang}, J.-M. 2023, \apj, 942, 112

\bibitem[{{Edelson} {et~al.}(2015){Edelson}, {Gelbord}, {Horne}, {McHardy},
  {Peterson}, {Ar{\'e}valo}, {Breeveld}, {De Rosa}, {Evans}, {Goad}, {Kriss},
  {Brandt}, {Gehrels}, {Grupe}, {Kennea}, {Kochanek}, {Nousek}, {Papadakis},
  {Siegel}, {Starkey}, {Uttley}, {Vaughan}, {Young}, {Barth}, {Bentz},
  {Brewer}, {Crenshaw}, {Dalla Bont{\`a}}, {De Lorenzo-C{\'a}ceres}, {Denney},
  {Dietrich}, {Ely}, {Fausnaugh}, {Grier}, {Hall}, {Kaastra}, {Kelly},
  {Korista}, {Lira}, {Mathur}, {Netzer}, {Pancoast}, {Pei}, {Pogge},
  {Schimoia}, {Treu}, {Vestergaard}, {Villforth}, {Yan}, \& {Zu}}]{edelson2015}
{Edelson}, R., {Gelbord}, J.~M., {Horne}, K., {et~al.} 2015, \apj, 806, 129

\bibitem[{{Elvis} {et~al.}(2002){Elvis}, {Marengo}, \& {Karovska}}]{elvis2002}
{Elvis}, M., {Marengo}, M., \& {Karovska}, M. 2002, \apjl, 567, L107

\bibitem[{{Fausnaugh} {et~al.}(2016){Fausnaugh}, {Denney}, {Barth}, {Bentz},
  {Bottorff}, {Carini}, {Croxall}, {De Rosa}, {Goad}, {Horne}, {Joner},
  {Kaspi}, {Kim}, {Klimanov}, {Kochanek}, {Leonard}, {Netzer}, {Peterson},
  {Schn{\"u}lle}, {Sergeev}, {Vestergaard}, {Zheng}, {Zu}, {Anderson},
  {Ar{\'e}valo}, {Bazhaw}, {Borman}, {Boroson}, {Brandt}, {Breeveld}, {Brewer},
  {Cackett}, {Crenshaw}, {Dalla Bont{\`a}}, {De Lorenzo-C{\'a}ceres},
  {Dietrich}, {Edelson}, {Efimova}, {Ely}, {Evans}, {Filippenko}, {Flatland},
  {Gehrels}, {Geier}, {Gelbord}, {Gonzalez}, {Gorjian}, {Grier}, {Grupe},
  {Hall}, {Hicks}, {Horenstein}, {Hutchison}, {Im}, {Jensen}, {Jones},
  {Kaastra}, {Kelly}, {Kennea}, {Kim}, {Korista}, {Kriss}, {Lee}, {Lira},
  {MacInnis}, {Manne-Nicholas}, {Mathur}, {McHardy}, {Montouri}, {Musso},
  {Nazarov}, {Norris}, {Nousek}, {Okhmat}, {Pancoast}, {Papadakis}, {Parks},
  {Pei}, {Pogge}, {Pott}, {Rafter}, {Rix}, {Saylor}, {Schimoia}, {Siegel},
  {Spencer}, {Starkey}, {Sung}, {Teems}, {Treu}, {Turner}, {Uttley},
  {Villforth}, {Weiss}, {Woo}, {Yan}, \& {Young}}]{Fausnaugh2016}
{Fausnaugh}, M.~M., {Denney}, K.~D., {Barth}, A.~J., {et~al.} 2016, \apj, 821,
  56

\bibitem[{{Ferland} {et~al.}(2017){Ferland}, {Chatzikos}, {Guzm{\'a}n},
  {Lykins}, {van Hoof}, {Williams}, {Abel}, {Badnell}, {Keenan}, {Porter}, \&
  {Stancil}}]{2017RMxAA..53..385F}
{Ferland}, G.~J., {Chatzikos}, M., {Guzm{\'a}n}, F., {et~al.} 2017, \rmxaa, 53,
  385

\bibitem[{{Garc{\'\i}a-Bernete} {et~al.}(2022){Garc{\'\i}a-Bernete},
  {Rigopoulou}, {Alonso-Herrero}, {Pereira-Santaella}, {Roche}, \&
  {Kerkeni}}]{garcia_bernete2022}
{Garc{\'\i}a-Bernete}, I., {Rigopoulou}, D., {Alonso-Herrero}, A., {et~al.}
  2022, \mnras, 509, 4256

\bibitem[{{Garnica} {et~al.}(2022){Garnica}, {Negrete}, {Marziani}, {Dultzin},
  {{\'S}niegowska}, \& {Panda}}]{Garnica_etal_2022}
{Garnica}, K., {Negrete}, C.~A., {Marziani}, P., {et~al.} 2022, \aap, 667, A105

\bibitem[{{Gaskell}(2017)}]{gaskell2017}
{Gaskell}, C.~M. 2017, \mnras, 467, 226

\bibitem[{{Gaskell} {et~al.}(2023){Gaskell}, {Anderson}, {Birmingham}, \&
  {Ghosh}}]{gasell2023}
{Gaskell}, C.~M., {Anderson}, F.~C., {Birmingham}, S.~{\'A}., \& {Ghosh}, S.
  2023, \mnras, 519, 4082

\bibitem[{{Gaskell} {et~al.}(2007){Gaskell}, {Klimek}, \&
  {Nazarova}}]{gaskell2007}
{Gaskell}, C.~M., {Klimek}, E.~S., \& {Nazarova}, L.~S. 2007, arXiv e-prints,
  arXiv:0711.1025

\bibitem[{{Goad} \& {Koratkar}(1998)}]{1998ApJ...495..718G}
{Goad}, M. \& {Koratkar}, A. 1998, \apj, 495, 718

\bibitem[{{Goad} {et~al.}(2016){Goad}, {Korista}, {De Rosa}, {Kriss},
  {Edelson}, {Barth}, {Ferland}, {Kochanek}, {Netzer}, {Peterson}, {Bentz},
  {Bisogni}, {Crenshaw}, {Denney}, {Ely}, {Fausnaugh}, {Grier}, {Gupta},
  {Horne}, {Kaastra}, {Pancoast}, {Pei}, {Pogge}, {Skielboe}, {Starkey},
  {Vestergaard}, {Zu}, {Anderson}, {Ar{\'e}valo}, {Bazhaw}, {Borman},
  {Boroson}, {Bottorff}, {Brandt}, {Breeveld}, {Brewer}, {Cackett}, {Carini},
  {Croxall}, {Dalla Bont{\`a}}, {De Lorenzo-C{\'a}ceres}, {Dietrich},
  {Efimova}, {Evans}, {Filippenko}, {Flatland}, {Gehrels}, {Geier}, {Gelbord},
  {Gonzalez}, {Gorjian}, {Grupe}, {Hall}, {Hicks}, {Horenstein}, {Hutchison},
  {Im}, {Jensen}, {Joner}, {Jones}, {Kaspi}, {Kelly}, {Kennea}, {Kim}, {Kim},
  {Klimanov}, {Lee}, {Leonard}, {Lira}, {MacInnis}, {Manne-Nicholas}, {Mathur},
  {McHardy}, {Montouri}, {Musso}, {Nazarov}, {Norris}, {Nousek}, {Okhmat},
  {Papadakis}, {Parks}, {Pott}, {Rafter}, {Rix}, {Saylor}, {Schimoia},
  {Schn{\"u}lle}, {Sergeev}, {Siegel}, {Spencer}, {Sung}, {Teems}, {Treu},
  {Turner}, {Uttley}, {Villforth}, {Weiss}, {Woo}, {Yan}, {Young}, \&
  {Zheng}}]{goad2016}
{Goad}, M.~R., {Korista}, K.~T., {De Rosa}, G., {et~al.} 2016, \apj, 824, 11

\bibitem[{{Goodrich}(1995)}]{Goodrich_1995}
{Goodrich}, R.~W. 1995, \apj, 440, 141

\bibitem[{{Gravity Collaboration} {et~al.}(2023){Gravity Collaboration},
  {Amorim}, {Bourdarot}, {Brandner}, {Cao}, {Cl{\'e}net}, {Davies}, {de Zeeuw},
  {Dexter}, {Drescher}, {Eckart}, {Eisenhauer}, {Fabricius}, {F{\"o}rster
  Schreiber}, {Garcia}, {Genzel}, {Gillessen}, {Gratadour}, {H{\"o}nig},
  {Kishimoto}, {Lacour}, {Lutz}, {Millour}, {Netzer}, {Ott}, {Paumard},
  {Perraut}, {Perrin}, {Peterson}, {Petrucci}, {Pfuhl}, {Prieto}, {Rouan},
  {Santos}, {Shangguan}, {Shimizu}, {Sternberg}, {Straubmeier}, {Sturm},
  {Tacconi}, {Tristram}, {Widmann}, \& {Woillez}}]{GRAVITY_2023}
{Gravity Collaboration}, {Amorim}, A., {Bourdarot}, G., {et~al.} 2023, \aap,
  669, A14

\bibitem[{{Guhathakurta} \& {Draine}(1989)}]{guhathakurta1989}
{Guhathakurta}, P. \& {Draine}, B.~T. 1989, \apj, 345, 230

\bibitem[{{Guise} {et~al.}(2022){Guise}, {H{\"o}nig}, {Gorjian}, {Barth},
  {Almeyda}, {Pei}, {Cenko}, {Edelson}, {Filippenko}, {Joner}, {Laney}, {Li},
  {Malkan}, {Nguyen}, \& {Zheng}}]{Guise_etal_2022}
{Guise}, E., {H{\"o}nig}, S.~F., {Gorjian}, V., {et~al.} 2022, \mnras, 516,
  4898

\bibitem[{{Hickox} \& {Alexander}(2018)}]{Hickox_Alexander_2018}
{Hickox}, R.~C. \& {Alexander}, D.~M. 2018, \araa, 56, 625

\bibitem[{{H{\"o}nig}(2019)}]{seba2019}
{H{\"o}nig}, S.~F. 2019, \apj, 884, 171

\bibitem[{{H{\"o}nig} \& {Kishimoto}(2017)}]{seba2017}
{H{\"o}nig}, S.~F. \& {Kishimoto}, M. 2017, \apjl, 838, L20

\bibitem[{{H{\"o}nig} {et~al.}(2012){H{\"o}nig}, {Kishimoto}, {Antonucci},
  {Marconi}, {Prieto}, {Tristram}, \& {Weigelt}}]{Honig_etal_2012}
{H{\"o}nig}, S.~F., {Kishimoto}, M., {Antonucci}, R., {et~al.} 2012, \apj, 755,
  149

\bibitem[{{H{\"o}nig} {et~al.}(2010){H{\"o}nig}, {Kishimoto}, {Gandhi},
  {Smette}, {Asmus}, {Duschl}, {Polletta}, \& {Weigelt}}]{Honig_etal_2010}
{H{\"o}nig}, S.~F., {Kishimoto}, M., {Gandhi}, P., {et~al.} 2010, \aap, 515,
  A23

\bibitem[{{H{\"o}nig} {et~al.}(2013){H{\"o}nig}, {Kishimoto}, {Tristram},
  {Prieto}, {Gandhi}, {Asmus}, {Antonucci}, {Burtscher}, {Duschl}, \&
  {Weigelt}}]{seba2013}
{H{\"o}nig}, S.~F., {Kishimoto}, M., {Tristram}, K.~R.~W., {et~al.} 2013, \apj,
  771, 87

\bibitem[{{Horne} {et~al.}(2021){Horne}, {De Rosa}, {Peterson}, {Barth}, {Ely},
  {Fausnaugh}, {Kriss}, {Pei}, {Bentz}, {Cackett}, {Edelson}, {Eracleous},
  {Goad}, {Grier}, {Kaastra}, {Kochanek}, {Krongold}, {Mathur}, {Netzer},
  {Proga}, {Tejos}, {Vestergaard}, {Villforth}, {Adams}, {Anderson},
  {Ar{\'e}valo}, {Beatty}, {Bennert}, {Bigley}, {Bisogni}, {Borman}, {Boroson},
  {Bottorff}, {Brandt}, {Breeveld}, {Brotherton}, {Brown}, {Brown}, {Canalizo},
  {Carini}, {Clubb}, {Comerford}, {Corsini}, {Crenshaw}, {Croft}, {Croxall},
  {Dalla Bont{\`a}}, {Deason}, {Dehghanian}, {De Lorenzo-C{\'a}ceres},
  {Denney}, {Dietrich}, {Done}, {Efimova}, {Evans}, {Ferland}, {Filippenko},
  {Flatland}, {Fox}, {Gardner}, {Gates}, {Gehrels}, {Geier}, {Gelbord},
  {Gonzalez}, {Gorjian}, {Greene}, {Grupe}, {Gupta}, {Hall}, {Henderson},
  {Hicks}, {Holmbeck}, {Holoien}, {Hutchison}, {Im}, {Jensen}, {Johnson},
  {Joner}, {Jones}, {Kaspi}, {Kelly}, {Kennea}, {Kim}, {Kim}, {Kim}, {King},
  {Klimanov}, {Korista}, {Lau}, {Lee}, {Leonard}, {Li}, {Lira}, {Lochhaas},
  {Ma}, {MacInnis}, {Malkan}, {Manne-Nicholas}, {Mauerhan}, {McGurk},
  {McHardy}, {Montuori}, {Morelli}, {Mosquera}, {Mudd},
  {M{\"u}ller-S{\'a}nchez}, {Nazarov}, {Norris}, {Nousek}, {Nguyen}, {Ochner},
  {Okhmat}, {Pancoast}, {Papadakis}, {Parks}, {Penny}, {Pizzella}, {Pogge},
  {Poleski}, {Pott}, {Rafter}, {Rix}, {Runnoe}, {Saylor}, {Schimoia},
  {Schn{\"u}lle}, {Scott}, {Sergeev}, {Shappee}, {Shivvers}, {Siegel},
  {Simonian}, {Siviero}, {Skielboe}, {Somers}, {Spencer}, {Starkey}, {Stevens},
  {Sung}, {Tayar}, {Treu}, {Turner}, {Uttley}, {Van Saders}, {Vican},
  {Villanueva}, {Weiss}, {Woo}, {Yan}, {Young}, {Yuk}, {Zheng}, {Zhu}, \&
  {Zu}}]{Horne_etal_2021}
{Horne}, K., {De Rosa}, G., {Peterson}, B.~M., {et~al.} 2021, \apj, 907, 76

\bibitem[{{Jaiswal} {et~al.}(2023){Jaiswal}, {Prince}, {Panda}, \&
  {Czerny}}]{vikram2023}
{Jaiswal}, V.~K., {Prince}, R., {Panda}, S., \& {Czerny}, B. 2023, \aap, 670,
  A147

\bibitem[{{Khadka} {et~al.}(2021){Khadka}, {Yu}, {Zaja{\v{c}}ek},
  {Martinez-Aldama}, {Czerny}, \& {Ratra}}]{khadka2021}
{Khadka}, N., {Yu}, Z., {Zaja{\v{c}}ek}, M., {et~al.} 2021, \mnras, 508, 4722

\bibitem[{{Kirkpatrick} {et~al.}(2012){Kirkpatrick}, {Pope}, {Alexander},
  {Charmandaris}, {Daddi}, {Dickinson}, {Elbaz}, {Gabor}, {Hwang}, {Ivison},
  {Mullaney}, {Pannella}, {Scott}, {Altieri}, {Aussel}, {Bournaud}, {Buat},
  {Coia}, {Dannerbauer}, {Dasyra}, {Kartaltepe}, {Leiton}, {Lin}, {Magdis},
  {Magnelli}, {Morrison}, {Popesso}, \& {Valtchanov}}]{kirkpatrick2012}
{Kirkpatrick}, A., {Pope}, A., {Alexander}, D.~M., {et~al.} 2012, \apj, 759,
  139

\bibitem[{{Kishimoto} {et~al.}(2011){Kishimoto}, {H{\"o}nig}, {Antonucci},
  {Barvainis}, {Kotani}, {Tristram}, {Weigelt}, \&
  {Levin}}]{Kishimoto_etal_2011}
{Kishimoto}, M., {H{\"o}nig}, S.~F., {Antonucci}, R., {et~al.} 2011, \aap, 527,
  A121

\bibitem[{{Kishimoto} {et~al.}(2007){Kishimoto}, {H{\"o}nig}, {Beckert}, \&
  {Weigelt}}]{kishimoto2007}
{Kishimoto}, M., {H{\"o}nig}, S.~F., {Beckert}, T., \& {Weigelt}, G. 2007,
  \aap, 476, 713

\bibitem[{{Korista} \& {Goad}(2000)}]{Korista_Goad_2000}
{Korista}, K.~T. \& {Goad}, M.~R. 2000, \apj, 536, 284

\bibitem[{{Korista} \& {Goad}(2019)}]{2019MNRAS.489.5284K}
{Korista}, K.~T. \& {Goad}, M.~R. 2019, \mnras, 489, 5284

\bibitem[{{Koshida} {et~al.}(2014){Koshida}, {Minezaki}, {Yoshii}, {Kobayashi},
  {Sakata}, {Sugawara}, {Enya}, {Suganuma}, {Tomita}, {Aoki}, \&
  {Peterson}}]{koshida2014}
{Koshida}, S., {Minezaki}, T., {Yoshii}, Y., {et~al.} 2014, \apj, 788, 159

\bibitem[{{Kriss} {et~al.}(2019){Kriss}, {De Rosa}, {Ely}, {Peterson},
  {Kaastra}, {Mehdipour}, {Ferland}, {Dehghanian}, {Mathur}, {Edelson},
  {Korista}, {Arav}, {Barth}, {Bentz}, {Brandt}, {Crenshaw}, {Dalla Bont{\`a}},
  {Denney}, {Done}, {Eracleous}, {Fausnaugh}, {Gardner}, {Goad}, {Grier},
  {Horne}, {Kochanek}, {McHardy}, {Netzer}, {Pancoast}, {Pei}, {Pogge},
  {Proga}, {Silva}, {Tejos}, {Vestergaard}, {Adams}, {Anderson}, {Ar{\'e}valo},
  {Beatty}, {Behar}, {Bennert}, {Bianchi}, {Bigley}, {Bisogni},
  {Boissay-Malaquin}, {Borman}, {Bottorff}, {Breeveld}, {Brotherton}, {Brown},
  {Brown}, {Cackett}, {Canalizo}, {Cappi}, {Carini}, {Clubb}, {Comerford},
  {Coker}, {Corsini}, {Costantini}, {Croft}, {Croxall}, {Deason}, {De
  Lorenzo-C{\'a}ceres}, {De Marco}, {Dietrich}, {Di Gesu}, {Ebrero}, {Evans},
  {Filippenko}, {Flatland}, {Gates}, {Gehrels}, {Geier}, {Gelbord}, {Gonzalez},
  {Gorjian}, {Grupe}, {Gupta}, {Hall}, {Henderson}, {Hicks}, {Holmbeck},
  {Holoien}, {Hutchison}, {Im}, {Jensen}, {Johnson}, {Joner}, {Kaspi}, {Kelly},
  {Kelly}, {Kennea}, {Kim}, {Kim}, {Kim}, {King}, {Klimanov}, {Krongold},
  {Lau}, {Lee}, {Leonard}, {Li}, {Lira}, {Lochhaas}, {Ma}, {MacInnis},
  {Malkan}, {Manne-Nicholas}, {Matt}, {Mauerhan}, {McGurk}, {Montuori},
  {Morelli}, {Mosquera}, {Mudd}, {M{\"u}ller-S{\'a}nchez}, {Nazarov}, {Norris},
  {Nousek}, {Nguyen}, {Ochner}, {Okhmat}, {Paltani}, {Parks}, {Pinto},
  {Pizzella}, {Poleski}, {Ponti}, {Pott}, {Rafter}, {Rix}, {Runnoe}, {Saylor},
  {Schimoia}, {Schn{\"u}lle}, {Scott}, {Sergeev}, {Shappee}, {Shivvers},
  {Siegel}, {Simonian}, {Siviero}, {Skielboe}, {Somers}, {Spencer}, {Starkey},
  {Stevens}, {Sung}, {Tayar}, {Teems}, {Treu}, {Turner}, {Uttley}, {. Van
  Saders}, {Vican}, {Villforth}, {Villanueva}, {Walton}, {Waters}, {Weiss},
  {Woo}, {Yan}, {Yuk}, {Zheng}, {Zhu}, \& {Zu}}]{kriss2019}
{Kriss}, G.~A., {De Rosa}, G., {Ely}, J., {et~al.} 2019, \apj, 881, 153

\bibitem[{{Krolik} {et~al.}(1981){Krolik}, {McKee}, \& {Tarter}}]{krolik1981}
{Krolik}, J.~H., {McKee}, C.~F., \& {Tarter}, C.~B. 1981, \apj, 249, 422

\bibitem[{{Lawther} {et~al.}(2018){Lawther}, {Goad}, {Korista}, {Ulrich}, \&
  {Vestergaard}}]{lawther2018}
{Lawther}, D., {Goad}, M.~R., {Korista}, K.~T., {Ulrich}, O., \& {Vestergaard},
  M. 2018, \mnras, 481, 533

\bibitem[{{Leftley} {et~al.}(2021){Leftley}, {Tristram}, {H{\"o}nig}, {Asmus},
  {Kishimoto}, \& {Gandhi}}]{Leftley_etal_2021}
{Leftley}, J.~H., {Tristram}, K. R.~W., {H{\"o}nig}, S.~F., {et~al.} 2021,
  \apj, 912, 96

\bibitem[{{L{\'o}pez-Gonzaga} {et~al.}(2016){L{\'o}pez-Gonzaga}, {Burtscher},
  {Tristram}, {Meisenheimer}, \& {Schartmann}}]{Lopez-Gonzaga_etal_2016}
{L{\'o}pez-Gonzaga}, N., {Burtscher}, L., {Tristram}, K.~R.~W., {Meisenheimer},
  K., \& {Schartmann}, M. 2016, \aap, 591, A47

\bibitem[{{Lyu} \& {Rieke}(2021)}]{Lyu_Rieke_2021}
{Lyu}, J. \& {Rieke}, G.~H. 2021, \apj, 912, 126

\bibitem[{{Lyu} {et~al.}(2019){Lyu}, {Rieke}, \& {Smith}}]{Lyu_etal_2019}
{Lyu}, J., {Rieke}, G.~H., \& {Smith}, P.~S. 2019, \apj, 886, 33

\bibitem[{{Maiolino} {et~al.}(2001){Maiolino}, {Salvati}, {Marconi}, \&
  {Antonucci}}]{Maiolino_etal_2001}
{Maiolino}, R., {Salvati}, M., {Marconi}, A., \& {Antonucci}, R.~R.~J. 2001,
  \aap, 375, 25

\bibitem[{{Mao} {et~al.}(2018){Mao}, {Kaastra}, {Mehdipour}, {Gu},
  {Costantini}, {Kriss}, {Bianchi}, {Branduardi-Raymont}, {Behar}, {Di Gesu},
  {Ponti}, {Petrucci}, \& {Ebrero}}]{mao2018}
{Mao}, J., {Kaastra}, J.~S., {Mehdipour}, M., {et~al.} 2018, \aap, 612, A18

\bibitem[{{Martin} \& {Ferland}(1980)}]{martin1980}
{Martin}, P.~G. \& {Ferland}, G.~J. 1980, \apjl, 235, L125

\bibitem[{{Mart{\'\i}nez-Aldama} {et~al.}(2019){Mart{\'\i}nez-Aldama},
  {Czerny}, {Kawka}, {Karas}, {Panda}, {Zaja{\v{c}}ek}, \&
  {{\.Z}ycki}}]{MartinezAldama2019}
{Mart{\'\i}nez-Aldama}, M.~L., {Czerny}, B., {Kawka}, D., {et~al.} 2019, \apj,
  883, 170

\bibitem[{{Marziani} {et~al.}(2021){Marziani}, {Berton}, {Panda}, \&
  {Bon}}]{Marziani_etal_2021}
{Marziani}, P., {Berton}, M., {Panda}, S., \& {Bon}, E. 2021, Universe, 7, 484

\bibitem[{{Marziani} {et~al.}(2023){Marziani}, {Panda}, {Deconto Machado}, \&
  {Del Olmo}}]{Marziani_etal_2023}
{Marziani}, P., {Panda}, S., {Deconto Machado}, A., \& {Del Olmo}, A. 2023,
  Galaxies, 11, 52

\bibitem[{{Mehdipour} {et~al.}(2016){Mehdipour}, {Kaastra}, {Kriss}, {Cappi},
  {Petrucci}, {De Marco}, {Ponti}, {Steenbrugge}, {Behar}, {Bianchi},
  {Branduardi-Raymont}, {Costantini}, {Ebrero}, {Di Gesu}, {Matt}, {Paltani},
  {Peterson}, {Ursini}, \& {Whewell}}]{mehdipour2016}
{Mehdipour}, M., {Kaastra}, J.~S., {Kriss}, G.~A., {et~al.} 2016, \aap, 588,
  A139

\bibitem[{{Mehdipour} {et~al.}(2015){Mehdipour}, {Kaastra}, {Kriss}, {Cappi},
  {Petrucci}, {Steenbrugge}, {Arav}, {Behar}, {Bianchi}, {Boissay},
  {Branduardi-Raymont}, {Costantini}, {Ebrero}, {Di Gesu}, {Harrison}, {Kaspi},
  {De Marco}, {Matt}, {Paltani}, {Peterson}, {Ponti}, {Pozo Nu{\~n}ez}, {De
  Rosa}, {Ursini}, {de Vries}, {Walton}, \& {Whewell}}]{2015A&A...575A..22M}
{Mehdipour}, M., {Kaastra}, J.~S., {Kriss}, G.~A., {et~al.} 2015, \aap, 575,
  A22

\bibitem[{{Mor} \& {Trakhtenbrot}(2011)}]{Mor_Trakhtenbrot_2011}
{Mor}, R. \& {Trakhtenbrot}, B. 2011, \apjl, 737, L36

\bibitem[{{Naddaf} \& {Czerny}(2022)}]{naddaf2022}
{Naddaf}, M.~H. \& {Czerny}, B. 2022, \aap, 663, A77

\bibitem[{{Naddaf} {et~al.}(2021){Naddaf}, {Czerny}, \&
  {Szczerba}}]{naddaf2021}
{Naddaf}, M.-H., {Czerny}, B., \& {Szczerba}, R. 2021, \apj, 920, 30

\bibitem[{{Naddaf} {et~al.}(2022){Naddaf}, {Martinez-Aldama}, {Marziani},
  {Panda}, {Sniegowska}, \& {Czerny}}]{naddaf_BAL_2022}
{Naddaf}, M.~H., {Martinez-Aldama}, M.~L., {Marziani}, P., {et~al.} 2022, arXiv
  e-prints, arXiv:2212.08222

\bibitem[{{Nenkova} {et~al.}(2008){Nenkova}, {Sirocky}, {Nikutta},
  {Ivezi{\'c}}, \& {Elitzur}}]{nenkova2008}
{Nenkova}, M., {Sirocky}, M.~M., {Nikutta}, R., {Ivezi{\'c}}, {\v{Z}}., \&
  {Elitzur}, M. 2008, \apj, 685, 160

\bibitem[{{Netzer}(2022)}]{netzr2022}
{Netzer}, H. 2022, \mnras, 509, 2637

\bibitem[{{Netzer} \& {Laor}(1993)}]{Netzer_Laor_1993}
{Netzer}, H. \& {Laor}, A. 1993, \apjl, 404, L51

\bibitem[{{Netzer} \& {Marziani}(2010)}]{netzer2010}
{Netzer}, H. \& {Marziani}, P. 2010, \apj, 724, 318

\bibitem[{{Neugebauer} {et~al.}(1979){Neugebauer}, {Oke}, {Becklin}, \&
  {Matthews}}]{neugebauer1979}
{Neugebauer}, G., {Oke}, J.~B., {Becklin}, E.~E., \& {Matthews}, K. 1979, \apj,
  230, 79

\bibitem[{{Osterbrock}(1981)}]{osterbrock1981}
{Osterbrock}, D.~E. 1981, \apj, 249, 462

\bibitem[{{Osterbrock} \& {Ferland}(2006)}]{osterbrock_book_2006}
{Osterbrock}, D.~E. \& {Ferland}, G.~J. 2006, {Astrophysics of gaseous nebulae
  and active galactic nuclei}

\bibitem[{{Panda}(2021{\natexlab{a}})}]{Panda_PhD_2021}
{Panda}, S. 2021{\natexlab{a}}, PhD thesis, Polish Academy of Sciences,
  Institute of Physics

\bibitem[{{Panda}(2021{\natexlab{b}})}]{Panda_2021}
{Panda}, S. 2021{\natexlab{b}}, \aap, 650, A154

\bibitem[{{Panda}(2022)}]{Panda_2022}
{Panda}, S. 2022, Frontiers in Astronomy and Space Sciences, 9, 850409

\bibitem[{{Panda} {et~al.}(2022){Panda}, {Bon}, {Marziani}, \&
  {Bon}}]{Panda_etal_2022}
{Panda}, S., {Bon}, E., {Marziani}, P., \& {Bon}, N. 2022, Astronomische
  Nachrichten, 343, e210091

\bibitem[{{Panda} {et~al.}(2018){Panda}, {Czerny}, {Adhikari}, {Hryniewicz},
  {Wildy}, {Kuraszkiewicz}, \& {{\'S}niegowska}}]{Panda_etal_2018}
{Panda}, S., {Czerny}, B., {Adhikari}, T.~P., {et~al.} 2018, \apj, 866, 115

\bibitem[{{Panda} {et~al.}(2020){Panda}, {Mart{\'\i}nez-Aldama}, {Marinello},
  {Czerny}, {Marziani}, \& {Dultzin}}]{Panda_etal_2020}
{Panda}, S., {Mart{\'\i}nez-Aldama}, M.~L., {Marinello}, M., {et~al.} 2020,
  \apj, 902, 76

\bibitem[{{Panda} {et~al.}(2019{\natexlab{a}}){Panda}, {Mart{\'\i}nez-Aldama},
  \& {Zaja{\v{c}}ek}}]{2019FrASS...6...75P}
{Panda}, S., {Mart{\'\i}nez-Aldama}, M.~L., \& {Zaja{\v{c}}ek}, M.
  2019{\natexlab{a}}, Frontiers in Astronomy and Space Sciences, 6, 75

\bibitem[{{Panda} {et~al.}(2019{\natexlab{b}}){Panda}, {Marziani}, \&
  {Czerny}}]{Panda_etal_2019}
{Panda}, S., {Marziani}, P., \& {Czerny}, B. 2019{\natexlab{b}}, \apj, 882, 79

\bibitem[{{Pei} {et~al.}(2017){Pei}, {Fausnaugh}, {Barth}, {Peterson}, {Bentz},
  {De Rosa}, {Denney}, {Goad}, {Kochanek}, {Korista}, {Kriss}, {Pogge},
  {Bennert}, {Brotherton}, {Clubb}, {Dalla Bont{\`a}}, {Filippenko}, {Greene},
  {Grier}, {Vestergaard}, {Zheng}, {Adams}, {Beatty}, {Bigley}, {Brown},
  {Brown}, {Canalizo}, {Comerford}, {Coker}, {Corsini}, {Croft}, {Croxall},
  {Deason}, {Eracleous}, {Fox}, {Gates}, {Henderson}, {Holmbeck}, {Holoien},
  {Jensen}, {Johnson}, {Kelly}, {Kim}, {King}, {Lau}, {Li}, {Lochhaas}, {Ma},
  {Manne-Nicholas}, {Mauerhan}, {Malkan}, {McGurk}, {Morelli}, {Mosquera},
  {Mudd}, {Muller Sanchez}, {Nguyen}, {Ochner}, {Ou-Yang}, {Pancoast}, {Penny},
  {Pizzella}, {Poleski}, {Runnoe}, {Scott}, {Schimoia}, {Shappee}, {Shivvers},
  {Simonian}, {Siviero}, {Somers}, {Stevens}, {Strauss}, {Tayar}, {Tejos},
  {Treu}, {Van Saders}, {Vican}, {Villanueva}, {Yuk}, {Zakamska}, {Zhu},
  {Anderson}, {Ar{\'e}valo}, {Bazhaw}, {Bisogni}, {Borman}, {Bottorff},
  {Brandt}, {Breeveld}, {Cackett}, {Carini}, {Crenshaw}, {De
  Lorenzo-C{\'a}ceres}, {Dietrich}, {Edelson}, {Efimova}, {Ely}, {Evans},
  {Ferland}, {Flatland}, {Gehrels}, {Geier}, {Gelbord}, {Grupe}, {Gupta},
  {Hall}, {Hicks}, {Horenstein}, {Horne}, {Hutchison}, {Im}, {Joner}, {Jones},
  {Kaastra}, {Kaspi}, {Kelly}, {Kennea}, {Kim}, {Kim}, {Klimanov}, {Lee},
  {Leonard}, {Lira}, {MacInnis}, {Mathur}, {McHardy}, {Montouri}, {Musso},
  {Nazarov}, {Netzer}, {Norris}, {Nousek}, {Okhmat}, {Papadakis}, {Parks},
  {Pott}, {Rafter}, {Rix}, {Saylor}, {Schn{\"u}lle}, {Sergeev}, {Siegel},
  {Skielboe}, {Spencer}, {Starkey}, {Sung}, {Teems}, {Turner}, {Uttley},
  {Villforth}, {Weiss}, {Woo}, {Yan}, {Young}, \& {Zu}}]{pei2017}
{Pei}, L., {Fausnaugh}, M.~M., {Barth}, A.~J., {et~al.} 2017, \apj, 837, 131

\bibitem[{{Peterson}(1987)}]{peterson1987}
{Peterson}, B.~M. 1987, \apj, 312, 79

\bibitem[{{Peterson} {et~al.}(1991){Peterson}, {Balonek}, {Barker}, {Bechtold},
  {Bertram}, {Bochkarev}, {Bolte}, {Bond}, {Boroson}, {Carini}, {Carone},
  {Christensen}, {Clements}, {Cochran}, {Cohen}, {Crampton}, {Dietrich},
  {Elvis}, {Ferguson}, {Filippenko}, {Fricke}, {Gaskell}, {Halpern}, {Huchra},
  {Hutchings}, {Kollatschny}, {Koratkar}, {Korista}, {Krolik}, {Lame}, {Laor},
  {Leacock}, {MacAlpine}, {Malkan}, {Maoz}, {Miller}, {Morris}, {Netzer},
  {Oliveira}, {Penfold}, {Penston}, {Perez}, {Pogge}, {Richmond}, {Romanishin},
  {Rosenblatt}, {Saddlemyer}, {Sadun}, {Sawyer}, {Shields}, {Shapovalova},
  {Smith}, {Smith}, {Smith}, {Sun}, {Thiele}, {Turner}, {Veilleux}, {Wagner},
  {Weymann}, {Wilkes}, {Wills}, {Wills}, \& {Younger}}]{peterson1991}
{Peterson}, B.~M., {Balonek}, T.~J., {Barker}, E.~S., {et~al.} 1991, \apj, 368,
  119

\bibitem[{{Peterson} {et~al.}(2002){Peterson}, {Berlind}, {Bertram},
  {Bischoff}, {Bochkarev}, {Borisov}, {Burenkov}, {Calkins}, {Carrasco},
  {Chavushyan}, {Chornock}, {Dietrich}, {Doroshenko}, {Ezhkova}, {Filippenko},
  {Gilbert}, {Huchra}, {Kollatschny}, {Leonard}, {Li}, {Lyuty}, {Malkov},
  {Matheson}, {Merkulova}, {Mikhailov}, {Modjaz}, {Onken}, {Pogge}, {Pronik},
  {Qian}, {Romano}, {Sergeev}, {Sergeeva}, {Shapovalova}, {Spiridonova}, {Tao},
  {Tokarz}, {Valdes}, {Vlasiuk}, {Wagner}, \& {Wilkes}}]{2002ApJ...581..197P}
{Peterson}, B.~M., {Berlind}, P., {Bertram}, R., {et~al.} 2002, \apj, 581, 197

\bibitem[{{Pozo Nu{\~n}ez} {et~al.}(2023){Pozo Nu{\~n}ez}, {Bruckmann},
  {Deesamutara}, {Czerny}, {Panda}, {Lobban}, {Pietrzy{\'n}ski}, \&
  {Polsterer}}]{Pozo-Nunez_etal_2023}
{Pozo Nu{\~n}ez}, F., {Bruckmann}, C., {Deesamutara}, S., {et~al.} 2023,
  \mnras, 522, 2002

\bibitem[{{Pozo Nu{\~n}ez} {et~al.}(2015){Pozo Nu{\~n}ez}, {Ramolla},
  {Westhues}, {Haas}, {Chini}, {Steenbrugge}, {Barr Dom{\'\i}nguez},
  {Kaderhandt}, {Hackstein}, {Kollatschny}, {Zetzl}, {Hodapp}, \&
  {Murphy}}]{Pozo-Nunez_etal_2015}
{Pozo Nu{\~n}ez}, F., {Ramolla}, M., {Westhues}, C., {et~al.} 2015, \aap, 576,
  A73

\bibitem[{{Ramos Almeida} \& {Ricci}(2017)}]{ramos_almeida_2017}
{Ramos Almeida}, C. \& {Ricci}, C. 2017, Nature Astronomy, 1, 679

\bibitem[{{Rokaki} {et~al.}(1993){Rokaki}, {Collin-Souffrin}, \&
  {Magnan}}]{rokaki1993}
{Rokaki}, E., {Collin-Souffrin}, S., \& {Magnan}, C. 1993, \aap, 272, 8

\bibitem[{{R{\'o}{\.z}a{\'n}ska}(1999)}]{agata1999}
{R{\'o}{\.z}a{\'n}ska}, A. 1999, \mnras, 308, 751

\bibitem[{{R{\'o}{\.z}a{\'n}ska} {et~al.}(2006){R{\'o}{\.z}a{\'n}ska},
  {Goosmann}, {Dumont}, \& {Czerny}}]{agata2006}
{R{\'o}{\.z}a{\'n}ska}, A., {Goosmann}, R., {Dumont}, A.~M., \& {Czerny}, B.
  2006, \aap, 452, 1

\bibitem[{{R{\'o}{\.z}a{\'n}ska} {et~al.}(2017){R{\'o}{\.z}a{\'n}ska},
  {Kunneriath}, {Czerny}, {Adhikari}, \& {Karas}}]{agata2017}
{R{\'o}{\.z}a{\'n}ska}, A., {Kunneriath}, D., {Czerny}, B., {Adhikari}, T.~P.,
  \& {Karas}, V. 2017, \mnras, 464, 2090

\bibitem[{{Rudy} \& {Puetter}(1982)}]{rudy1982}
{Rudy}, R.~J. \& {Puetter}, R.~C. 1982, \apj, 263, 43

\bibitem[{{Schn{\"u}lle} {et~al.}(2015){Schn{\"u}lle}, {Pott}, {Rix},
  {Peterson}, {De Rosa}, \& {Shappee}}]{Schnulle_etal_2015}
{Schn{\"u}lle}, K., {Pott}, J.~U., {Rix}, H.~W., {et~al.} 2015, \aap, 578, A57

\bibitem[{{Sergeev} {et~al.}(2007){Sergeev}, {Doroshenko}, {Dzyuba},
  {Peterson}, {Pogge}, \& {Pronik}}]{Sergeev_etal_2007}
{Sergeev}, S.~G., {Doroshenko}, V.~T., {Dzyuba}, S.~A., {et~al.} 2007, \apj,
  668, 708

\bibitem[{{Shablovinskaya} {et~al.}(2020){Shablovinskaya}, {Afanasiev}, \&
  {Popovi{\'c}}}]{Shablovinskaya_etal_2020}
{Shablovinskaya}, E.~S., {Afanasiev}, V.~L., \& {Popovi{\'c}}, L.~{\v{c}}.
  2020, \apj, 892, 118

\bibitem[{{Shapovalova} {et~al.}(2004){Shapovalova}, {Doroshenko}, {Bochkarev},
  {Burenkov}, {Carrasco}, {Chavushyan}, {Collin}, {Vald{\'e}s}, {Borisov},
  {Dumont}, {Vlasuyk}, {Chilingarian}, {Fioktistova}, \&
  {Martinez}}]{shapovalova2004}
{Shapovalova}, A.~I., {Doroshenko}, V.~T., {Bochkarev}, N.~G., {et~al.} 2004,
  \aap, 422, 925

\bibitem[{{Siebenmorgen} {et~al.}(2015){Siebenmorgen}, {Heymann}, \&
  {Efstathiou}}]{siebenmorgen2015}
{Siebenmorgen}, R., {Heymann}, F., \& {Efstathiou}, A. 2015, \aap, 583, A120

\bibitem[{{{\'S}niegowska} {et~al.}(2021){{\'S}niegowska}, {Marziani},
  {Czerny}, {Panda}, {Mart{\'\i}nez-Aldama}, {del Olmo}, \&
  {D'Onofrio}}]{Sniegowska_etal_2021}
{{\'S}niegowska}, M., {Marziani}, P., {Czerny}, B., {et~al.} 2021, \apj, 910,
  115

\bibitem[{{Sobrino Figaredo} {et~al.}(2020){Sobrino Figaredo}, {Haas},
  {Ramolla}, {Chini}, {Blex}, {Hodapp}, {Murphy}, {Kollatschny}, {Chelouche},
  \& {Kaspi}}]{Sobrino_Figaredo_etal_2020}
{Sobrino Figaredo}, C., {Haas}, M., {Ramolla}, M., {et~al.} 2020, \aj, 159, 259

\bibitem[{{Stalevski} {et~al.}(2019){Stalevski}, {Tristram}, \&
  {Asmus}}]{Stalevski_etal_2019}
{Stalevski}, M., {Tristram}, K. R.~W., \& {Asmus}, D. 2019, \mnras, 484, 3334

\bibitem[{{Temple} {et~al.}(2021){Temple}, {Banerji}, {Hewett}, {Rankine}, \&
  {Richards}}]{temple2021}
{Temple}, M.~J., {Banerji}, M., {Hewett}, P.~C., {Rankine}, A.~L., \&
  {Richards}, G.~T. 2021, \mnras, 501, 3061

\bibitem[{{Tristram} {et~al.}(2014){Tristram}, {Burtscher}, {Jaffe},
  {Meisenheimer}, {H{\"o}nig}, {Kishimoto}, {Schartmann}, \&
  {Weigelt}}]{Tristram_etal_2014}
{Tristram}, K. R.~W., {Burtscher}, L., {Jaffe}, W., {et~al.} 2014, \aap, 563,
  A82

\bibitem[{{van Hoof} {et~al.}(2004){van Hoof}, {Weingartner}, {Martin}, {Volk},
  \& {Ferland}}]{2004MNRAS.350.1330V}
{van Hoof}, P.~A.~M., {Weingartner}, J.~C., {Martin}, P.~G., {Volk}, K., \&
  {Ferland}, G.~J. 2004, \mnras, 350, 1330

\bibitem[{{Weingartner} {et~al.}(2006){Weingartner}, {Draine}, \&
  {Barr}}]{2006ApJ...645.1188W}
{Weingartner}, J.~C., {Draine}, B.~T., \& {Barr}, D.~K. 2006, \apj, 645, 1188

\bibitem[{{Wildy} {et~al.}(2021){Wildy}, {Landt}, {Ward}, {Czerny}, \&
  {Kynoch}}]{wildy2021}
{Wildy}, C., {Landt}, H., {Ward}, M.~J., {Czerny}, B., \& {Kynoch}, D. 2021,
  \mnras, 500, 2063

\bibitem[{{Yang} {et~al.}(2020){Yang}, {Shen}, {Liu}, {Aguena}, {Annis},
  {Avila}, {Banerji}, {Bertin}, {Brooks}, {Burke}, {Carnero Rosell}, {Carrasco
  Kind}, {da Costa}, {De Vicente}, {Desai}, {Diehl}, {Doel}, {Flaugher},
  {Fosalba}, {Frieman}, {Garcia-Bellido}, {Gerdes}, {Gruen}, {Gruendl},
  {Gschwend}, {Gutierrez}, {Hinton}, {Hollowood}, {Honscheid}, {Kuropatkin},
  {Maia}, {March}, {Marshall}, {Martini}, {Melchior}, {Menanteau}, {Miquel},
  {Paz-Chinchon}, {Malag{\'o}n}, {Romer}, {Sanchez}, {Scarpine}, {Schubnell},
  {Serrano}, {Sevilla}, {Smith}, {Suchyta}, {Tarle}, {Varga}, \&
  {Wilkinson}}]{Yang_etal_2020}
{Yang}, Q., {Shen}, Y., {Liu}, X., {et~al.} 2020, \apj, 900, 58

\bibitem[{{Yu} {et~al.}(2023){Yu}, {Martini}, {Penton}, {Davis}, {Kochanek},
  {Lewis}, {Lidman}, {Malik}, {Sharp}, {Tucker}, {Aguena}, {Annis}, {Bertin},
  {Bocquet}, {Brooks}, {Carnero Rosell}, {Carollo}, {Carrasco Kind},
  {Carretero}, {Costanzi}, {da Costa}, {Pereira}, {De Vicente}, {Diehl},
  {Doel}, {Everett}, {Ferrero}, {Garc{\'\i}a-Bellido}, {Gatti}, {Gerdes},
  {Gruen}, {Gruendl}, {Gschwend}, {Gutierrez}, {Hinton}, {Hollowood},
  {Honscheid}, {James}, {Kuehn}, {Mena-Fern{\'a}ndez}, {Menanteau}, {Miquel},
  {Nichol}, {Paz-Chinch{\'o}n}, {Pieres}, {Plazas Malag{\'o}n}, {Raveri},
  {Romer}, {Sanchez}, {Scarpine}, {Sevilla-Noarbe}, {Smith}, {Suchyta},
  {Swanson}, {Tarle}, {Vincenzi}, {Walker}, \& {Weaverdyck}}]{Yu2023}
{Yu}, Z., {Martini}, P., {Penton}, A., {et~al.} 2023, \mnras, 522, 4132

\bibitem[{{Zaja{\v{c}}ek} {et~al.}(2020){Zaja{\v{c}}ek}, {Czerny},
  {Martinez-Aldama}, {Ra{\l}owski}, {Olejak}, {Panda}, {Hryniewicz},
  {{\'S}niegowska}, {Naddaf}, {Pych}, {Pietrzy{\'n}ski}, {Sobrino Figaredo},
  {Haas}, {{\'S}redzi{\'n}ska}, {Krupa}, {Kurcz}, {Udalski}, {Gorski}, \&
  {Sarna}}]{zajacek2020}
{Zaja{\v{c}}ek}, M., {Czerny}, B., {Martinez-Aldama}, M.~L., {et~al.} 2020,
  \apj, 896, 146

\bibitem[{{Zaja{\v{c}}ek} {et~al.}(2021){Zaja{\v{c}}ek}, {Czerny},
  {Martinez-Aldama}, {Ra{\l}owski}, {Olejak}, {Przy{\l}uski}, {Panda},
  {Hryniewicz}, {{\'S}niegowska}, {Naddaf}, {Prince}, {Pych},
  {Pietrzy{\'n}ski}, {Sobrino Figaredo}, {Haas}, {{\'S}redzi{\'n}ska}, {Krupa},
  {Kurcz}, {Udalski}, {Karas}, {Sarna}, {Worters}, {Sefako}, \&
  {Genade}}]{zajacek2021}
{Zaja{\v{c}}ek}, M., {Czerny}, B., {Martinez-Aldama}, M.~L., {et~al.} 2021,
  \apj, 912, 10

\end{thebibliography}

\end{document}